\documentclass[aps,prd,preprintnumbers,twocolumn,showpacs,nofootinbib]{revtex4}
\usepackage[dvips]{graphicx}
\usepackage{enumerate}
\usepackage{amsmath,amssymb}
\usepackage{mathrsfs}
\usepackage[dvips]{graphicx}
\usepackage{bm}

\begin{document}
\preprint{CHIBA-EP-230v4, 2018.07.14}

\title{Gauge-independent Brout-Englert-Higgs mechanism and Yang-Mills theory with a gauge-invariant gluon mass term}

\author{Kei-Ichi Kondo$^{1}$}
\email{kondok@faculty.chiba-u.jp}

\affiliation{$^1$Department of Physics,  
Graduate School of Science, 
Chiba University, Chiba 263-8522, Japan
}
\begin{abstract}
For the Yang-Mills theory coupled to a single scalar field in the fundamental representation of the gauge group, we present a gauge-independent description of the Brout-Englert-Higgs mechanism by which massless gauge bosons acquire their mass.  The new  description should be compared with the conventional gauge-dependent description relying on the spontaneous gauge symmetry breaking due to a choice of the non-vanishing vacuum expectation value of the scalar field.
In this paper we focus our consideration on the fundamental scalar field which extends the previous work done for the Yang-Mills theory with an adjoint scalar field.
Moreover, we show that the Yang-Mills theory with a gauge-invariant mass term is obtained from the corresponding gauge-scalar model when the radial degree of freedom (length) of the scalar field is fixed. 
The result obtained in this paper is regarded as a continuum realization of the Fradkin-Shenker continuity and Osterwalder-Seiler theorem for the complementarity between Higgs regime and Confinement regime which was given in the gauge-invariant framework of the lattice gauge theory. 
Moreover, we discuss how confinement is investigated through the gauge-independent Brout-Englert-Higgs mechanism by starting with the complementary gauge-scalar model.

\end{abstract}

\pacs{12.38.Aw, 21.65.Qr}

\maketitle

\section{Introduction}

In a previous paper \cite{Kondo16} we have proposed a gauge-independent description  of the Brout-Englert-Higgs (BEH) or Higgs mechanism \cite{Higgs1,Higgs2,Higgs3} which is defined to be a mechanism for  massless gauge bosons to acquire their mass.  In discussing the BEH mechanism we exclude the Higgs particle from our consideration to focus on the mass generation for the gauge boson. 
The conventional description of the BEH mechanism states that massless gauge bosons become massive vector bosons by absorbing the Nambu-Goldstone particles \cite{NJL61,Goldstone61} associated with the spontaneous breaking of the gauge symmetry.  
This description requires a non-vanishing vacuum expectation value of the scalar field, 
\begin{align}
\langle 0| \phi(x) |0 \rangle=v ,
\nonumber
\end{align}
which is clearly gauge dependent and impossible to be realized without fixing the gauge due to the Elitzur theorem \cite{Elitzur75} (see also Introduction of \cite{Kondo16} for more details).  
In the new description \cite{Kondo16}, instead, the scalar field is supposed to obey a  gauge-invariant condition which forces the radial length of the scalar field to have a certain fixed value $v (v>0)$, 
\begin{align}
||\phi(x)||=v \quad (||\phi(x)||^2=v^2) ,
\nonumber
\end{align}
without breaking the gauge symmetry. 
This enables us to extract the massive vector modes in the operator level from the original Yang-Mills field in a  gauge-independent manner. 
Consequently, we can introduce a gauge-invariant mass term   in the pure Yang-Mills theory.
Such an example was already given in the case of an adjoint scalar field for a gauge group $SU(2)$  in a previous work \cite{Kondo16}.
 In this paper, we extend the gauge-independent description of the BEH mechanism to include a fundamental scalar field.

 In the gauge-scalar model with a fundamental scalar field, the Higgs phase and the Confinement phase are analytically continued in the phase diagram and are not separated by the thermodynamic phase transition . 
This fact is called the Fradkin--Shenker continuity \cite{FS79} which was derived in the gauge-invariant framework of lattice gauge theory for the gauge-scalar model with a radially fixed fundamental scalar field and is understood as a special case of the analytic continuity due to the Osterwalder--Seiler theorem \cite{OS78}. 
The Fradkin--Shenker continuity holds for various compact groups (continuous and discrete), e.g., $G=SU(N)$, $U(1)$, $Z(N)$.
This leads to the idea of ``\textbf{complementarity}'' between Higgs and Confinement. 
In the gauge-scalar model with an adjoint scalar field, on the other hand, the Higgs and the Confinement phases are not analytically continued  in the phase diagram and are distinct phases separated by the phase transition. 
Such phase structures of the  gauge-scalar models were also confirmed by numerical simulations on the lattice. See \cite{lattice-gauge-scalar-fund} for a fundamental scalar and \cite{lattice-gauge-scalar-adj} for an adjoint scalar case. 
(The Fradkin--Shenker continuity does not hold even for the fundamental scalar model if the scalar field has the variable length with a sufficiently small $0 < \lambda \ll 1$ self-interaction coupling $\lambda$ for the potential term $V=\lambda (||\phi(x)||^2-v^2)^2$ \cite{gauge-scalar-rad-var}. Higgs phase and Confinement phase are not analytically continued in the phase diagram and separated by the  phase transition.)
Therefore, the result obtained in this paper can be regarded as an explicit realization of the Fradkin--Shenker continuity in the framework of the continuum field theory. 
Consequently, this work enables one to investigate confinement and mass gap in the pure Yang-Mills theory as the implications of the BEH mechanism in the gauge-scalar model \cite{FMS80,tHooft80}. See e.g. \cite{Maas17} for a recent review on the BEH mechanism on the lattice.

Due to the gauge-invariant description of the BEH mechanism, we can extract the massive modes $\mathscr{W}_\mu$ from the original gauge field $\mathscr{A}_\mu$ in the gauge-independent way. 
The massive  vector mode $\mathscr{W}_\mu$ will rapidly fall off in the distance and hence it is identified with the short-distance (or high-energy) mode. 
Therefore,  massive vector modes $\mathscr{W}$ mediate only the short-range force between quark sources. 
Consequently, there must exist the nontrivial residual mode, 
$
\mathscr{V}_\mu=\mathscr{A}_\mu-\mathscr{W}_\mu
$
 which is identified with the long-distance (or low-energy) mode. 
In the Confinement phase, the residual mode will mediate the long-range force which is responsible for quark confinement in the sense of area law falloff of the Wilson loop average or linear  quark potential. 
In other words, the new description of the BEH mechanism allows one to decompose the original gauge field $\mathscr{A}$ into the massive vector mode $\mathscr{W}$ and the residual gauge mode $
\mathscr{V}$, 
\begin{align}
\mathscr{A}=\mathscr{W}+\mathscr{V} .
\nonumber
\end{align}
In the case of the fundamental scalar field, there are no massless gauge bosons in the residual mode $\mathscr{V}$ once the BEH mechanism occurs.  In fact, we show that the residual gauge mode $\mathscr{V}$ has exactly the same form as the pure gauge $\mathscr{V}=ig^{-1}UdU^{-1}$ with the group element $U$ which is written in terms of the scalar field $\Phi$ alone. 
Therefore the residual gauge mode is trivial in the perturbative treatment.  However, the residual gauge mode can be nontrivial in the non-perturbative treatment.  Indeed, solitons and defects \cite{soliton} converging to the pure gauge in the long distance can be good candidates for the dominant components responsible for confinement. 
This situation should be compared with the case of the adjoint scalar field where the residual gauge mode include massless gauge boson which is able to mediate the long-range force \cite{Polyakov77b,tHP74}.


Moreover, the new description of the BEH mechanism provides physical meaning of the gauge-covariant field decomposition obtained by Cho, Duan-Ge and Faddeev--Niemi (CDGFN) \cite{Cho80,DG79,FN99,KMS06,KMS05,Kondo06} in the pure Yang-Mills theory, and  allows various decompositions \cite{KSM08} other than the original CDGFN one by considering all possible complementary gauge-scalar models. See e.g. \cite{KKSS15} for a review. 

Last but not least,
I must mention the preceding works related to the present work.  
The fact itself that the mass term of the gauge field can be written in a gauge-invariant form has long been known since the St\"uckelberg formalism \cite{Stueckelberg38} for the $U(1)$ symmetry case.
In fact, the non-Abelian generalization has been done half a century ago by Kunimasa and Goto \cite{KG67}, Slavnov and Faddeev \cite{SF70}, Cornwall \cite{Cornwall74,Cornwall82} and the others \cite{DT86}, see e.g., \cite{DTT88,RRA04} for reviews. 
A locally gauge-invariant description of gluon mass demands the existence of massless scalar fields which do not appear in the $S$ matrix.  Although the massless scalar fields looks like the Nambu-Goldstone counterparts in spontaneously broken gauge theories, there is no spontaneous symmetry breaking associated with gluon mass generation according to  \cite{Cornwall74,Cornwall82}.

One of the purposes of this paper is to realize a gauge-invariant gluon mass term of the Yang-Mills theory taking always the connection with confinement problem. 
The gauge-invariant gluon mass is introduced through a gauge-independent BEH mechanism by extending the Yang-Mills theory (in the covariant gauge) to the gauge-invariant gauge-scalar model in which Confinement phase is expected to be analytically continued to the Higgs phase in the sense of the Fradkin-Shenker continuity.
The massive Yang-Mills theory obtained in this way can be efficient for understanding the confining decoupling solution in the Landau gauge which was confirmed by the numerical simulations on the lattice \cite{decoupling-lattice} and was examined in the analytical approach \cite{decoupling-analytical}, see e.g. the proceedings \cite{QCD-TNT09,Ghent-QCD} for the related works. 
This enables one to provide a novel explanation for the Cornwall claim that the gluon mass can be dynamically generated in the gauge-invariant way without spontaneous symmetry breaking. 
In this paper the gluon mass is described by a locally gauge-invariant mass term as if the mass were a constant. 
Because such a mass term is enough to reproduce the decoupling solution in the low-momentum region below a few GeV with gluon confinement \cite{Bowmanetal07},  as demonstrated explicitly in a subsequent paper \cite{Kondo-etal18}.

In the complementary gauge-scalar model, the scalar field $\Phi$ and the gauge field $\mathscr{A}$ are not independent field variables, because we intend to obtain the massive pure Yang-Mills theory which does not contain the scalar field $\Phi$.
Therefore, the scalar field $\Phi$  which corresponds to the St\"uckelberg field in the preceding works is to be eliminated in favor of the gauge field $\mathscr{A}$.  
This is in principle achieved by solving the constraint called the reduction condition as an off-shell equation, which is different from solving the equation of motion for the scalar field $\Phi$ \cite{Cornwall82,DT86}. 
However, the resulting expression for the scalar field $\Phi$ would be given by a complicated form, e.g., an infinite series in perturbation theory. 
In particular, the scalar field $\Phi$ becomes trivial when the gauge fields is transverse, namely, in the Landau gauge $\partial^\mu \mathscr{A}_\mu=0$.  
Consequently, the resulting massive Yang-Mills theory in the covariant gauge-fixing term and the associated Faddeev-Popov ghost term becomes power-counting renormalizable in the perturbative framework, as will be shown in \cite{Kondo-etal18}. 
Moreover, the entire theory is invariant under the  Becchi-Rouet-Stora-Tyutin (BRST) transformation.  The nilpotency of the BRST transformations ensures the unitarity of the theory in the physical subspace of the total state vector space determined by zero BRST charge according to Kugo and Ojima \cite{KO79}. 
In view of these, we recall that the Curci-Ferrari model \cite{CF76b} which is not invariant under the ordinary BRST transformation can be made invariant under the modified BRST transformation.  However, this fact does not guarantee the unitarity due to the lack of usual nilpotency of the modified BRST transformation, see e.g., \cite{Kondo13}.


This paper is organized as follows. 
In section II, we discuss the case of the Abelian gauge group $U(1)$ before attacking the non-Abelian gauge group. 
In section III, we treat the gauge group $SU(2)$. 
In section IV, we give the construction of the color direction field which is useful to discuss magnetic monopoles in Yang-Mills theory to compare the fundamental scalar case with the adjoint scalar case. 
The final section is devoted to conclusion and discussion. 
In Appendix A, we give an example of the exact solution for the topological soliton for the $U(1)$ gauge-scalar model with a radially fixed scalar field. 
In Appendix B, we summarize the formulas for the $SU(2)$ gauge-scalar model in case of the fundamental and adjoint scalar fields for comparison.

\section{$U(1)$ gauge-scalar model}

\subsection{Radially fixed $U(1)$ gauge-scalar model}

The $U(1)$ gauge-scalar model is described by the Lagrangian density:
\begin{align}
\mathscr{L}_{\rm AH}(x) =& - \frac{1}{4} F_{\mu\nu}(x) F^{\mu\nu}(x) +  ( {D}_{\mu}[A]\phi(x))^{*} ({D}^{\mu}[A]\phi(x)) 
\nonumber\\& 
- V(\phi(x)) , 
\nonumber\\
V(\phi(x)) =& \frac{\lambda}{2} \left( \phi^{*}(x) \phi(x) - \frac{\mu^{2}}{\lambda} \right)^{2} , \ \phi(x) \in \mathbb{C} ,
\end{align}
where $F_{\mu\nu}(x)$ is the field strength for the $U(1)$ gauge field $A_{\mu}(x)$ given by $F_{\mu\nu}(x) = \partial_{\mu} A_{\nu}(x) - \partial_{\nu} A_{\mu}(x)$ and ${D}_{\mu}[A]$ is the covariant derivative defined by ${D}_{\mu}[A] = \partial_{\mu} - i q A_{\mu}(x)$ for the complex scalar field $\phi(x) \in \mathbb{C}$ with $q$ being the electric charge of $\phi(x)$. 
Here $*$ denotes the complex conjugate.

We focus on the $U(1)$ gauge-scalar model with a \textbf{radially fixed scalar field} which is described by the Lagrangian density:
\begin{align}
\mathscr{L}_{\rm RF}(x) =& - \frac{1}{4} F_{\mu\nu}(x) F^{\mu\nu}(x) +  ( {D}_{\mu}[A]\phi(x) )^{*}({D}^{\mu}[A]\phi(x)) 
\nonumber\\ &
+ u(x) \left(\phi^{*}(x) \phi(x) - \frac{1}{2} v^2 \right)  
 ,
\end{align}
where $u(x)$ is the \textbf{Lagrange multiplier field} to incorporate the gauge-invariant constraint that the radial degree of freedom, i.e., the absolute value of the scalar field is fixed, 
\begin{align}
\phi^{*}(x) \phi(x) - \frac{1}{2} v^2 =  0  
\Leftrightarrow 
|\phi(x)|= \frac{v}{\sqrt{2}} > 0. 
\label{U1-const}
\end{align}

\par
The $U(1)$ gauge-scalar model is a gauge theory with a local gauge invariance of the gauge group $U(1)$.
In fact, $\mathscr{L}_{\rm AH}$ is invariant under the $U(1)$ gauge transformation:
\begin{align}
\phi(x) \rightarrow \phi^{\prime}(x) =&  U(x) \phi(x), 
\quad  U(x)=e^{i q\chi (x)} \in U(1),
\nonumber\\  
A_{\mu}(x) \rightarrow A_{\mu}^{\prime}(x) 
=&  U(x)[A_{\mu}(x) + q^{-1} \partial_{\mu} ]U(x)^*  
\nonumber\\  
=&  A_{\mu}(x) + \partial_{\mu} \chi (x) . 
\end{align}
$\mathscr{L}_{\rm RF}$ is also invariant under the $U(1)$ gauge transformation, provided that $u(x)$ is gauge invariant.
Notice that the constraint (\ref{U1-const}) is a gauge-invariant condition.

\subsection{BEH mechanism for $U(1)$ gauge-scalar model}

We introduce the \textbf{normalized scalar field} $\hat{\phi}$ by
\begin{align}
 \hat{\phi}(x) := \phi(x)/\left(\frac{v}{\sqrt{2}}\right), \ v > 0   
.
\end{align} 
Then the normalized scalar field $\hat{\phi}$ is an element of the group $U(1)$:
\begin{align}
 \hat{\phi}(x) \in  G=U(1) 
,
\end{align}
since the above constraint (\ref{U1-const}) implies that the normalized scalar field $\hat{\phi}$ obeys the condition:
\begin{align}
\hat{\phi}^*(x) \hat{\phi}(x) = \hat{\phi}(x) \hat{\phi}^*(x) =  1 
.
\label{norm-U1}
\end{align} 

Then we introduce the vector boson field $W_\mu$ defined in terms of the the normalized scalar field $\hat{\phi}$ and the original gauge field $A_\mu$ as
\footnote{
By using $\hat{\phi}(x)  \hat{\phi}(x)^*=1$, we find
\begin{align}
 W_\mu(x) 
=&  iq^{-1}  \partial_{\mu} \hat{\phi}(x) \hat{\phi}(x)^* +  A_{\mu}(x) 
\nonumber\\
=& -iq^{-1} \hat{\phi}(x) \partial_{\mu} \hat{\phi}(x)^*  +  A_{\mu}(x)  
\label{W2-U1}
\end{align}
}
\begin{align}
	W_\mu(x) 
 :=& iq^{-1} ( {D}_{\mu}[A] \hat{\phi}(x)) \hat{\phi}(x)^* 
\nonumber\\
=& -iq^{-1} \hat{\phi}(x) ({D}_{\mu}[A] \hat{\phi}(x) )^*
\nonumber\\
=& \frac{1}{2} iq^{-1} [ ( {D}_{\mu}[A] \hat{\phi}(x))  \hat{\phi}(x)^* -  \hat{\phi}(x)	({D}_{\mu}[A]  \hat{\phi}(x) )^* ]
 .
\label{W1-U1}
\end{align}

We find that \textbf{the kinetic term of the scalar field $\phi$ is identical to the mass term of the vector boson field $W^\mu$ with the mass $M_W$}:
\begin{align}
 ( {D}_{\mu}[A] \phi(x))^* 	{D}^{\mu}[A] \phi(x) 
=& \frac{1}{2} M_W^2 W_\mu(x) W^\mu(x) , 
\nonumber\\  M_W :=& qv .
\end{align}
It is remarkable that the vector boson field $W^\mu$ is gauge-invariant and that \textbf{the mass term of the vector boson field  $W^\mu$ is gauge invariant}, as explicitly checked. 
This equivalence between the kinetic term and the mass term is easily shown by using (\ref{norm-U1}) and (\ref{W1-U1}) as  
\begin{align}
	 \left(\frac{v}{\sqrt{2}}\right)^2  q^2W_\mu W^\mu 
=& i (-i ) ( {D}_{\mu}[A] \phi ) \hat{\phi}^* \hat{\phi} 	({D}^{\mu}[A] \phi)^*  
\nonumber\\
=&  ( {D}_{\mu}[A] \phi )^* 	({D}^{\mu}[A] \phi) 
 .
\end{align}

To see the correspondence to the conventional viewpoint for the BEH mechanism, we replace the field $\phi$ by its vacuum expectation value,
\begin{align}
		\phi(x)   
\to \langle \phi(x) \rangle = \phi_\infty := \frac{v}{\sqrt{2}} e^{i\theta_\infty} 
 .
 \label{unitary-gauge-U1}
\end{align} 
Then the kinetic term reduces to the mass term:
\begin{align}
 & ( {D}_{\mu}[A] \phi(x))^* 	{D}^{\mu}[A] \phi(x)  
  \nonumber\\
\to&    [  iq  {\phi}_\infty^* A_{\mu}(x) ] 
 [  -iq A^{\mu}(x)  {\phi}_\infty ] 
=  \frac12 (qv)^2 A_{\mu}(x) 
   A^{\mu}(x)   .
\end{align}
In this replacement (\ref{unitary-gauge-U1}), $W_\mu$ reduces to the original gauge field,
\begin{align}
 W_\mu(x) \to  & iq^{-1} ( {D}_{\mu}[A(x)] \hat{\phi}_\infty) \hat{\phi}_\infty^* 
\nonumber\\
=& ig^{-1} ( -iq A_{\mu}(x) \hat{\phi}_\infty) \hat{\phi}_\infty^* = A_\mu(x) 
 ,
\end{align}
with the mass term 
\begin{align}
	 M_W^2 W_\mu(x) W^\mu(x) 
\to  M_W^2 A_\mu(x) A^\mu(x) .
\end{align}
Consequently, all components of the massless gauge boson become  massive.

In the conventional understanding of the BEH  mechanism, the original gauge symmetry $G$ is spontaneously broken completely with no residual gauge symmetry, which we call the \textbf{complete SSB}, $G= U(1) \to H=\{1 \}$.
The \textbf{Nambu-Goldstone mode} $\pi$ associated with the \textbf{spontaneous breaking of $U(1)$ symmetry} is identified as follows.
If we use  the representation: \textit{polar decomposition} for a radially fixed scalar field: 
\begin{align}
	&	\phi(x) = \frac{v}{\sqrt{2}} \hat{\phi}(x), \quad
\hat{\phi}(x) = e^{i \pi(x)/v} \in \mathbb{C} , \ \pi(x) \in \mathbb{R} ,
\end{align}
then the massive field $W_{\mu}$ is written as 
\begin{align}
W_{\mu}(x) = A_{\mu}(x) - M_W^{-1} \partial_{\mu} \pi(x) .
\end{align}
This is usually said that the Nambu-Goldstone mode $\pi$ associated with the spontaneous breaking of $U(1)$ symmetry is absorbed into the gauge field $A_\mu$ as the longitudinal mode to make the massive vector boson $W_\mu$. 

\textbf{The representation for $W_\mu$ given above (\ref{W1-U1}) is parameterization independent, namely, does not depend on the specific parameterization of the scalar field.} 
Therefore, we can use the other coordinate, e.g., 
\begin{align}
	&	\phi(x) = \frac{1}{\sqrt{2}} [v+\varphi(x) + i\chi(x)] .
\end{align}

\subsection{Field decomposition for $U(1)$ gauge-scalar model
}

The original gauge field $A_\mu$ is decomposed into the gauge-invariant massive vector field $W_\mu$ and the residual mode $V_\mu$:
\begin{align}
	A_\mu(x)  = W_\mu(x) + V_\mu(x) 
.
\label{A-decomposition-U1}
\end{align}
Under the gauge transformation $U(x) \in U(1)$, the original fields transform as
\begin{align}
 A_{\mu}(x) & \to  A_{\mu}^{\prime}(x) = A_{\mu}(x)  + iq^{-1} U(x) \partial_\mu U(x)^*  
 , \quad
\nonumber\\
 \phi(x) & \to \phi^{\prime}(x) = U(x) \phi(x) ,
\  U(x)=e^{i q\chi (x)} \in U(1).
\end{align}
As the massive vector field $W_\mu$ is gauge invariant,
\begin{align}
 W_\mu(x)   \to  W_{\mu}(x) 
 ,
 \label{W-gauge-transf}
\end{align}
while the residual field $V_\mu$ must transform like the original one $A_\mu$,
\begin{align}
 & V_{\mu}(x) \to  V_{\mu}^{\prime}(x) = V_{\mu}(x)  + iq^{-1} U(x) \partial_\mu U(x)^*  
.
\end{align}

To obtain the explicit expression for $V_\mu$, we observe  that $W_\mu=0$ is equivalent to the gauge-invariant condition for $V_\mu$  given by 
\begin{align}
	 {D}_{\mu}[V] \hat{\phi}(x) =  0 \Leftrightarrow 
	 \partial_{\mu} \hat{\phi}(x)   -iq V_{\mu}(x) \hat{\phi}(x)  = 0 
 .
\end{align}
The \textbf{residual mode} $V_\mu$ is obtained by solving this equation using $\hat{\phi}  \hat{\phi}^* = 1$ as
\begin{align}
 V_{\mu}(x) =&  -iq^{-1}\partial_{\mu} \hat{\phi}(x) \hat{\phi}(x)^*  =  iq^{-1} \hat{\phi}(x) \partial_{\mu}  \hat{\phi}(x)^* , 
\nonumber\\ &
 \hat{\phi}(x) \in  U(1) 
.
\label{V1-U1}
\end{align}
This agrees with the result of (\ref{W2-U1}).

In the replacement (\ref{unitary-gauge-U1}), the residual field $V_\mu$ vanishes (except the singular points),
\begin{align}
 V_{\mu}(x) \to  -iq^{-1}\partial_{\mu} \hat{\phi}_\infty \hat{\phi}^*_\infty  = 0 
 .
\end{align}

In the perturbative treatment the residual mode $V_\mu$ is trivial. But it is non-trivial in the non-perturbative treatment as discussed later.

\subsection{Reduction condition for $U(1)$ gauge theory}

In the $U(1)$ gauge-scalar model,  $A_\mu(x)$ and   $\phi(x)$ are independent field variables. 
However, the pure $U(1)$ gauge theory should be described by  $A_\mu(x)$ alone and hence  $\phi(x)$ must be supplied by the gauge field $A_\mu(x)$.
In other words, the scalar field $\phi(x)$ should be given as a functional of the gauge field $A_\mu(x)$.
This is achieved by imposing the appropriate constraint which we call the  \textbf{reduction condition}.

We proceed to find the reduction condition. 
\textbf{Imposing the reduction condition eliminates an extra degree of freedom introduced into the pure $U(1)$ gauge theory through the radially fixed complex scalar field $\hat{\phi} \in U(1)$, which is necessary to convert the $U(1)$ gauge-scalar theory to the pure $U(1)$ gauge theory.} 

To find the reduction condition, we consider the extended gauge theory with the enlarged gauge symmetry $U(1)_\omega \times U(1)_\theta$ according to  the procedure given in \cite{KMS06}. 
The infinitesimal form of the enlarged gauge transformation is given by
\begin{align}
\delta_{\omega,\theta} \phi(x) = iq \theta(x) \phi(x) ,  
\quad 
\delta_{\omega,\theta} A_{\mu}(x) =  \partial_\mu \omega(x) .  
\end{align}
Under the enlarged gauge transformation, $W_\mu$ transform as
\begin{align}
	\delta_{\omega,\theta} W_\mu(x) 
 =  -  \partial_{\mu} (\theta(x) - \omega(x) ) 
 ,
\label{dW}
\end{align}
because
\begin{align}
	&\delta_{\omega,\theta} W_\mu 
\nonumber\\
 =& iq^{-1} ( {D}_{\mu}[A] \delta_{\omega,\theta}  \hat{\phi} ) \hat{\phi} ^* 
+ iq^{-1} ( {D}_{\mu}[A] \hat{\phi} ) \delta_{\omega,\theta}  \hat{\phi} ^* 
\nonumber\\&
+ iq^{-1} ( -iq\delta_{\omega,\theta} A_\mu \hat{\phi} ) \hat{\phi} ^* 
\nonumber\\ 
 =& - ( {D}_{\mu}[A] (\theta \hat{\phi}) ) \hat{\phi} ^* 
+  ( {D}_{\mu}[A] \hat{\phi} )  \hat{\phi} ^* \theta
+     (\partial_{\mu}  \omega )  \hat{\phi}   \hat{\phi} ^* 
\nonumber\\ 
 =& - ( \partial_{\mu} (\theta \hat{\phi})-iqA_\mu  \theta \hat{\phi}) \hat{\phi} ^* 
+  ( \partial_{\mu} \hat{\phi} -iqA_\mu \hat{\phi})  \hat{\phi} ^* \theta
+ \partial_{\mu} \omega   
\nonumber\\ 
 =& - ( \partial_{\mu} (\theta \hat{\phi}) \hat{\phi}^* -iqA_\mu  \theta  )  
+  ( \partial_{\mu} \hat{\phi} \hat{\phi} ^* \theta -iqA_\mu     \theta )
+ \partial_{\mu}  \omega  
\nonumber\\ 
 =& -   \partial_{\mu} \theta -   \theta \partial_{\mu} \hat{\phi} \hat{\phi}^*   
+  \partial_{\mu} \hat{\phi} \hat{\phi} ^* \theta 
+ \partial_{\mu}  \omega
\nonumber\\ 
 =& -  \partial_{\mu} (\theta - \omega )
 .
\label{dW1}
\end{align}
Indeed, this transformation recovers the infinitesimal form of the original gauge transformation when $\theta=\omega$ (\ref{W-gauge-transf}):
\begin{align}
	\delta_{\omega} W_\mu(x) 
 = 0  
 .
\label{dW2}
\end{align}

Then the variation of the functional reads
\begin{align}
 \delta_{\omega,\theta} \int d^Dx \frac12 W_\mu W_\mu  
 =&	\int d^Dx W_\mu \delta_{\theta,\omega} W^\mu 
\nonumber\\ 
 =& \int d^Dx \left( -W_\mu \partial^{\mu} (\theta - \omega ) \right) 
\nonumber\\ 
 =& \int d^Dx (\theta - \omega ) (\partial^{\mu} W_\mu )    
,
\end{align}
where we have used the integration by parts in the third equality. 
Thus we obtain the reduction condition as a gauge-invariant condition:
\begin{align}
 \chi(x) := \partial^{\mu} W_\mu(x) = 0 
 .
\label{U1-reduction}
\end{align}
The reduction condition is rewritten in terms of the   scalar field $\hat{\phi}$ and the original gauge field $A_\mu$ as
\begin{align}
 \chi(x) :=& \partial^{\mu} [( {D}_{\mu}[A] \hat{\phi}(x)) \hat{\phi}(x)^*] = 0 
\nonumber\\
 \Longleftrightarrow 
 \chi(x) :=& - \partial^{\mu} [\hat{\phi}(x)	({D}_{\mu}[A] \hat{\phi}(x) )^*] = 0 
 .
\label{dW5}
\end{align}
The reduction condition must be gauge covariant equation and retain the same form under the gauge transformation. 
The obtained reduction condition (\ref{U1-reduction}) is actually gauge-invariant.

\subsection{Field equations to the reduction condition 
}

We discuss the relationship between the reduction condition and the field equation of the gauge-scalar model.

For the $U(1)$ gauge-scalar model with the quartic self-interacting potential, the field equations for $A_\mu$ and $\phi$  are given by
\begin{align}
0=\frac{\delta S_{\mathrm{AH}}}{\delta A^\mu(x)} 
=& \partial^\nu F_{\nu\mu}(x)  + iq [ \phi^*(x) D_\mu[A] \phi(x) 
\nonumber\\& 
-  (D_\mu[A] \phi(x))^* \phi(x) ]   ,
\nonumber\\
0=\frac{\delta S_{\mathrm{AH}}}{\delta \phi^*(x)} 
=& - D_\mu[A] D^\mu[A] \phi(x) 
\nonumber\\&
-  \lambda \left( \phi^*(x) \phi(x) - \frac{\mu^{2}}{\lambda} \right) \phi(x)   . 
\label{CP-AH-fe}
\end{align}

For the $U(1)$ gauge-scalar model with a radially fixed scalar field, the field equations for the fields $u$, $\phi$   and $A_\mu$ are respectively given by
\begin{align}
0= \frac{\delta S_{\rm RF}}{\delta u(x)} 
 =& \phi ^{*}(x)\phi(x) - \frac12 v^2   ,
 \label{eq-u1}
\\ 
0= \frac{\delta S_{\rm RF}}{\delta \phi^*(x)} 
 =& -  {D}_{\mu}[A] {D}^{\mu}[A] \phi(x) + \phi(x) u(x)   ,
 \label{eq-f1}
\\ 
0= \frac{\delta S_{\rm RF}}{\delta \phi (x)} 
 =& -  {D}_{\mu}[A]^* ({D}^{\mu}[A] \phi(x))^* +  u(x) \phi(x)^*    ,
 \label{eq-ff1}
\\ 
0= \frac{\delta S_{\rm RF}}{\delta A^\mu(x)} 
 =&  \partial^\nu F_{\nu\mu}(x) +   iq [({D}_{\mu}[A] \phi(x)) {\phi}(x)^*  
\nonumber\\&
-  {\phi}(x) ({D}_{\mu}[A] \phi(x))^* ]  
 ,
 \label{eq-A1}
\end{align}
where the  field equation (\ref{eq-A1}) for $A_\mu$ is equivalent to 
\begin{align}
0= \frac{\delta S_{\rm RF}}{\delta A^\mu(x)} 
 =& \partial^\nu F_{\nu\mu}(x) +   M_W^2 W_\mu(x)   .
 \label{eq-AAa}
\end{align}

We proceed to study the relationship between the reduction condition and the field equation. 
Due to (\ref{eq-u1}), the scalar field $\phi$ can be  normalized $\hat{\phi}$. 
Multiplying (\ref{eq-f1}) by $\hat{\phi}^*$   and (\ref{eq-ff1}) by $\hat{\phi}$ yields 
\begin{align}
 0 =& \{ -{D}_{\mu}[A]  ( {D}^{\mu}[A] \hat{\phi}) + \hat{\phi} u \} \hat{\phi}^*
\nonumber\\&
-  \hat{\phi} \{ -{D}_{\mu}[A]^* ({D}^{\mu}[A] \hat{\phi} )^* + u \hat{\phi}^* \} 
= 2iq \partial_\mu  W^\mu  .
 \label{eq-AA-U1}
\end{align}
Applying the derivative to (\ref{eq-A1}) or (\ref{eq-AAa}) yields 
\begin{align}
0 = \partial_\mu ( \partial_\nu  F^{\nu\mu} + M_W^2  W^\mu  ) = M_W^2 \partial_\mu  W^\mu 
 \label{eq-AA3}
 .
\end{align}
\textbf{If the fields $A$ and $\phi$ are a set of solutions of the field equations for the $U(1)$ gauge-scalar model with a radially fixed scalar field, they automatically satisfy the reduction condition (\ref{U1-reduction}) for pure $U(1)$ gauge theory.}

The conserved \textbf{Noether current} $J^\mu$ associated to the $U(1)$ global symmetry defined by
\begin{align}
\delta_{\theta} \phi(x) =  iq \theta \phi(x) , \  \delta_{\theta} \phi(x)^* =  -iq  \phi(x)^* \theta  , \ 
\delta_{\theta}  A_\mu(x) = 0, 
\end{align}
is given by 
\begin{align}
 J^\mu  =&  \theta^{-1} \left[   \frac{\partial \mathscr{L}}{\partial \partial_\mu \phi^*} \delta \phi^* + \delta \phi  \frac{\partial \mathscr{L}}{\partial \partial_\mu \phi}   \right]
\nonumber\\
=&    -i q ({D}_{\mu}[A] \phi) \phi ^* + i q\phi ({D}_{\mu}[A] \phi)^*  
.
 \label{Noether2}
\end{align}
Notice that 
$W^\mu$ is proportional to the Noether current $J^\mu$:
\begin{align}
 J^\mu  = - M_W^2  W^\mu  .
\label{Noether3}
\end{align}
Since the Noether current $J^\mu$ is conserved  $\partial_\mu J^\mu = 0$, the $W^\mu$ satisfies the (divergenceless) relation:
\begin{align}
 \partial_\mu W^\mu = 0 ,
\end{align}
This is identified with the subsidiary condition for the massive field $W^\mu$.

The conserved Noether charge becomes a generator of the $U(1)$ global transformation:
\begin{align}
 \delta \phi(x) = [ i\theta Q, \phi(x) ]
= i\theta \int d^dy [J^0(y), \phi(x) ]
= i \theta q\phi(x) ,
\end{align}
which is shown by using  
$
 J^0 = iq \phi \Pi_\phi - iq \phi^* \Pi_{\phi^*}
$
with $\Pi_\phi$ and $\Pi_{\phi^*}$ being the canonical momenta conjugate to $\phi$ and $\phi^*$ respectively. 
This is consistent with no SSB:
\begin{align}
 \langle 0| \delta \phi(x) |0 \rangle 
= i \theta q  \langle 0| \phi(x) |0 \rangle = 0 ,
\end{align}
since 
$\phi$ is a gauge non-invariant operator with vanishing vacuum expectation value.


\subsection{Topology for $U(1)$ gauge-scalar model
}

Notice that the residual field $V_{\mu}$ is of the pure gauge type.
The residual field can give the nonvanishing topological configurations.

The target space $\mathcal{M}$ of the scalar field (vacuum manifold) is $\mathcal{M}=U(1) \simeq S^1$. Therefore, we consider the map $\phi: S_\infty^n \to S^1$ from the $n$-dimensional sphere $S^n$ at infinity in the $D$-dimensional space-time to the vacuum manifold $U(1)$.  Then the  topological non-trivial configuration is characterized by the non-trivial homotopy group $\pi_n(U(1))=\pi_n(S^1)\not= 0$. 
The non-trivial homotopy is possible only when $n=1$, $\phi: S_\infty^1 \to S^1$, namely, the $U(1)$ field defined on the circle $S_\infty^1$ in the space-time with the non-trivial homotopy group $\pi_1(S^1) = \mathbb{Z}$. 
\footnote{
Notice that $\pi_n(S^1)=0$ for $n>1$, $\pi_n(S^n)=\mathbb{Z}$ and $\pi_n(S^m)=0$ for $m>n$.  
Incidentally, $m<n$ case is non-trivial in general, e.g., 
$\Pi_4(S^3) = \mathbb{Z}_2$.
}

The point-like defect such as monopoles arises if the vacuum manifold  $\mathcal{M}$ contains non-contractible two-surfaces like the sphere $S^2$. 
This occurs when the vacuum manifold $\mathcal{M}$ has a non-trivial second homotopy group $\pi_2(\mathcal{M}) \not= 0$. 
For this to occur, it suffice to know if the unbroken symmetry group $H$ has a non-trivial fundamental group $\pi_1(H) \not= 0$ assuming $\pi_1(G)=\pi_2(G)=0$.

The line-like defect such as vortex or string arises if the vacuum manifold $\mathcal{M}$ is not simply connected; that is, $\mathcal{M}$ contains enclosed holes about which loops can be trapped.  This topological property is revealed if the fundamental group of $\mathcal{M}$ is non-trivial, $\pi_1(\mathcal{M}) \not= 0$. 
The elements of $\pi_1(\mathcal{M})$ classify the different types of admissible solutions. 
For a connected and simply connected symmetry group $G$, the line-like defects can be classified by $\pi_0(H)$, the disconnected components of the unbroken subgroup $H$.

For example, the residual mode represents the \textbf{vortex solution} of the Nielsen-Olesen type for $D=2+1$ dimensions and instanton in $D=2$ dimensions. 
In the radially fixed case, we have the exact analytical solution as shown later in Appendix A.


\section{$SU(2)$ gauge-scalar model: fundamental scalar}

In this section we give a manifestly gauge-independent description of the BEH or Higgs mechanism for the $SU(2)$ Yang-Mills theory coupled to the scalar field in the fundamental representation. 
In the conventional description, the BEH mechanism of the $SU(2)$ gauge-scalar model is explained as a consequence of the complete spontaneous breaking of the original $SU(2)$ gauge symmetry.
The gauge-independent description to be given below does not rely on the spontaneous breaking of gauge symmetry. 

A typical example of an $SU(2)$ gauge-scalar model is  described by the Lagrangian density:%
\begin{align}
		\mathscr{L}_{\rm HK} =& -\frac{1}{2} {\rm tr}[ \mathscr{F} _{\mu\nu}(x)\mathscr{F}^{\mu\nu } (x)]
\nonumber\\ &
+ ({D}_{\mu}[\mathscr{A}]\Phi(x))^{\dagger} \cdot ({D}^{\mu}[\mathscr{A}]\Phi(x)) - V( \Phi(x) ), 
\nonumber\\ 
 V( \Phi(x) ) 
:=&  - \mu^2 \Phi(x)^{\dagger}  \cdot \Phi(x)
		+ \frac{\lambda}{2} (\Phi(x)^{\dagger}  \cdot \Phi(x))^2 
\nonumber\\ 
=& \frac{\lambda}{2}  \left( \Phi(x)^{\dagger}  \cdot \Phi(x) - \frac{\mu^2}{\lambda} \right)^2 + {\rm const.} , 
\nonumber\\ &
 \mu^2 \in \mathbb{R} , \ \lambda>0 .
\end{align}
We define the $SU(2)$ gauge field $\mathscr{A}_{\mu}$ by
\begin{align}
\mathscr{A}_{\mu}(x) = \mathscr{A}^A_{\mu}(x) T_A, \ T_A = \frac{1}{2} \sigma_A ,
\end{align}
its  field strength $\mathscr{F}_{\mu\nu} $  by
\begin{align}
\mathscr{F}_{\mu\nu} (x)=& \mathscr{F}^A_{\mu\nu} (x)T_A , 
\nonumber\\  
\mathscr{F}^A_{\mu\nu} (x) =&  \partial_\mu \mathscr{A}_\nu^A (x) -  \partial_\nu \mathscr{A}_\mu^A (x) + g \epsilon^{ABC} \mathscr{A}_\mu^B (x) \mathscr{A}_\nu^C (x),
\end{align}
and the covariant derivative ${D}_{\mu}$ in the fundamental representation by
\begin{equation}
		{D}_{\mu}[\mathscr{A}]  =  \partial_{\mu} -ig \mathscr{A}_{\mu}(x)     .
\end{equation}
Here $\Phi (x)$ is the \textbf{$SU(2)$ doublet} 
formed from two complex scalar  fields   $\bm{\phi}_1 (x), \bm{\phi}_2 (x)$ which are parameterized (by the reason clarified later) as
\begin{align}
		\Phi(x) =&
	 \begin{pmatrix}
			\bm{\phi}_1(x) \\ 
			\bm{\phi}_2(x)
	\end{pmatrix}   , \
 \bm{\phi}_1(x), \bm{\phi}_2(x) \in \mathbb{C}  
\nonumber\\  
=& \frac{1}{\sqrt{2}}
		 \begin{pmatrix}
			\phi_2(x) + i \phi_1(x) \\ 
			\phi_0(x) - i \phi_3(x)
	\end{pmatrix}  ,
\nonumber\\  &
  \ \phi_0(x), \phi_A(x) \in \mathbb{R} \ (A=1,2,3) .
\end{align} 

In what follows we focus on the $SU(2)$ gauge-fundamental scalar model with a \textbf{radially fixed scalar field} described by the Lagrangian density
\begin{align}
		\mathscr{L}_{\rm RF} =& -\frac{1}{2} {\rm tr}[ \mathscr{F} _{\mu\nu}(x)\mathscr{F}^{\mu\nu } (x)]
\nonumber\\& 
		+ ({D}_{\mu}[\mathscr{A}]\Phi(x))^{\dagger}  \cdot ({D}^{\mu}[\mathscr{A}]\Phi(x)) 
\nonumber\\& 
+ u(x) \left( \Phi(x)^{\dagger}  \cdot \Phi(x) - \frac{1}{2}v^2   \right) ,
\label{SU2-gauge-scalar-f1}
\end{align}
where $u(x)$ is the \textbf{Lagrange multiplier field} to incorporate the constraint that the radial degree of freedom or length of the scalar field is fixed $|\Phi(x)|=v/\sqrt{2}>0$: 
\begin{align}
		 \Phi(x)^{\dagger}  \cdot \Phi(x) - \frac{1}{2}v^2 = 0 .
\label{SU2-s-constraint1}
\end{align}

Both gauge-scalar models have the local $SU(2)$ gauge symmetry. 
Indeed, the Lagrangian density is invariant under the $SU(2)$ gauge transformation given by
\begin{align}
& \Phi(x)  \to \Phi'(x) =  U(x) \Phi(x) ,  
\nonumber\\
& \mathscr{A}_{\mu}(x)  \to \mathscr{A}_{\mu}'(x) =  U(x) \mathscr{A}_{\mu}(x) U(x)^{-1} + ig^{-1}  U(x) \partial_\mu U(x)^{-1} ,  
\nonumber\\ 
& U(x) =  e^{ig \omega (x) } \in SU(2) , \ \omega (x) := \omega^A(x) T_A , \ T_A = \frac{1}{2} \sigma_A ,
\end{align}
which has the infinitesimal version:
\begin{align}
		\delta_\omega \Phi(x) =& ig \omega (x) \Phi(x) , 
		\nonumber\\
		\delta_\omega \mathscr{A}_{\mu}(x) =& \mathscr{D}_\mu[ \mathscr{A}] \omega (x) , 
\end{align}
with the covariant derivative $\mathscr{D}_\mu[ \mathscr{A}]$  in the adjoint representation defined by 
\footnote{
In the usual convention, the covariant derivative acts differently on the fields transforming differently, so that there is no need to make distinction between $\mathscr{D}_\mu$ and $D_\mu$. 
In this paper, however, we adopt an unusual convention in which $D_\mu$ or $\mathscr{D}_\mu$ is respectively used when acting on the field in the fundamental  or adjoint representation to call attention. 
Notice that 
$
\mathscr{D}_\mu \Phi= D_\mu \Phi
$
and
$
\mathscr{D}_\mu \Phi^\dagger = ({D}_\mu \Phi)^\dagger
$.
}
\begin{align}
 \mathscr{D}_\mu[\mathscr{A}] := \partial_{\mu} -ig [\mathscr{A}_{\mu}(x), ~\cdot~ ]  .
\end{align}

In what follows, we consider the radially fixed scalar field satisfying the constraint:
\begin{align}
		 {\Phi}(x)^\dagger  \cdot {\Phi}(x) 
=& \bm{\phi}_1^*(x) \bm{\phi}_1(x) + \bm{\phi}_2^*(x) \bm{\phi}_2(x) 
\nonumber\\
=&  \frac{1}{2} (\phi_0^2(x)  + \phi_A(x) \phi_A(x)  ) =  \frac{1}{2}  v^2
.
\label{SU2-s-constraint2}
\end{align}
Due to the constraint, ${\Phi}(x)$ has three independent degrees of freedom. 
Notice that this constraint is gauge invariant.  
The Lagrange multiplier field $u(x)$ is supposed to be invariant under the $SU(2)$ gauge transformation.
Therefore, even after this constraint is imposed,  the gauge symmetry is left unbroken. 
The Higgs particle corresponds to the variable length degree of freedom of the scalar field. 
By imposing this constraint, therefore, we eliminate the Higgs particle mode to focus on the mass generation for the gauge boson alone, which facilitate discussing the relation of the gauge-scalar model to the pure Yang-Mills theory as shown below.

\subsection{Matrix scalar field}

We proceed to construct the gauge group element from the scalar field. 
For this purpose, we introduce the \textbf{matrix-valued scalar field} $\Theta$ by adding another $SU(2)$ doublet $\tilde\Phi:=i\tau_2 \Phi^*$ as
\begin{align}
		\Theta :=&
		  \begin{pmatrix} 
		\tilde\Phi &	\Phi 
	\end{pmatrix} 
=
		  \begin{pmatrix} 
		i\tau_2 \Phi^* &	\Phi 
	\end{pmatrix} 
=
		  \begin{pmatrix} 
		\bm{\phi}_2^* &	\bm{\phi}_1 \\ 
		-\bm{\phi}_1^* &	\bm{\phi}_2
	\end{pmatrix} 
\nonumber\\
=&  \frac{1}{\sqrt{2}} (\phi_0 {\boldsymbol 1} + i \phi_A \sigma^A) 
= \frac{1}{\sqrt{2}}
	 \begin{pmatrix}
			\phi_0 + i \phi_3 & \phi_2 + i \phi_1 \\ 
			- \phi_2+i \phi_1  & \phi_0 - i \phi_3
	\end{pmatrix}  
 ,
\nonumber\\ 
  i\tau_2 =& \epsilon 
= \begin{pmatrix} 
		0 &	1 \\ 
		-1 & 0
	\end{pmatrix}
 .
	\label{Theta1}
\end{align} 
Notice that $\tilde\Phi$ has the same gauge transformation property as $\Phi$. 
Then the matrix-valued scalar field $\Theta$ has the same gauge transformation as $\Phi$,
\begin{align}
  \Theta(x) \to \Theta^\prime(x) = U(x) \Theta(x)   , \ U(x) \in SU(2) 
 .
\end{align}

We find that ${\Theta}^\dagger {\Theta}$ and ${\Theta}{\Theta}^\dagger $ are proportional to the unit matrix $\bm{1} $:
\begin{align}
 {\Theta}(x)^\dagger  {\Theta}(x) =&  {\Theta}(x)  {\Theta}(x)^\dagger = \Phi(x)^\dagger  \cdot \Phi(x) \bm{1} 
\nonumber\\ 
 =& (|\bm{\phi}_1(x)|^2+|\bm{\phi}_2(x)|^2)\bm{1} 
 ,
\end{align} 
which is shown using 
\begin{align}
		\Theta^\dagger =&
		  \begin{pmatrix} 
		 \tilde\Phi^\dagger \\
	\Phi^\dagger 
	\end{pmatrix} 
=		  \begin{pmatrix} 
		\bm{\phi}_2 &	-\bm{\phi}_1 \\ 
		\bm{\phi}_1^* &	\bm{\phi}_2^*
	\end{pmatrix} 
=  \frac{1}{\sqrt{2}} (\phi_0 {\boldsymbol 1} - i \phi_A \sigma^A) 
\nonumber\\ 
=& \frac{1}{\sqrt{2}}
	 \begin{pmatrix}
			\phi_0 - i \phi_3 & - \phi_2 -i \phi_1  \\ 
			\phi_2 -i \phi_1  & \phi_0 + i \phi_3
	\end{pmatrix}  ,
	\label{Theta2}
\end{align}
where
\begin{align}
 \tilde\Phi^\dagger  
 =  (\epsilon \Phi^*)^\dagger
 = \Phi^t \epsilon^\dagger
 = \Phi^t \epsilon^t .
 \label{tilde-rel}
\end{align}

Then  we introduce the \textbf{normalized matrix-valued scalar field} $\hat{\Theta}$ by
\begin{align}
 \hat{\Theta}(x) =  {\Theta} (x)/\left(\frac{v}{\sqrt{2}}\right), \ v > 0   
.
\end{align} 
The above constraint (\ref{SU2-s-constraint1}) or (\ref{SU2-s-constraint2}) implies that the normalized scalar field $\hat{\Theta}$ obeys the conditions:
\begin{align}
\hat{\Theta}(x)^\dagger \hat{\Theta}(x) = \hat{\Theta}(x) \hat{\Theta}(x)^\dagger =  \bm{1} 
,
\label{norm}
\end{align} 
and
\begin{align}
 \det \hat{\Theta}(x) 
= 1 
.
\label{det}
\end{align}
Therefore, the normalized matrix-valued scalar field $\hat{\Theta}$ is an element of $SU(2)$:
\begin{align}
 \hat{\Theta}(x) \in G=SU(2) 
.
\end{align}
This is an important property to give a gauge-independent BEH mechanism later. 

The original kinetic term of the scalar field is rewritten in terms of the matrix-valued scalar field as 
\begin{align}
 ({D}_{\mu}[\mathscr{A}]\Phi)^{\dagger}  \cdot ({D}^{\mu}[\mathscr{A}]\Phi) 
= \frac{1}{2} {\rm tr}(  ( {D}_{\mu}[\mathscr{A}] \Theta(x))^\dagger 	{D}^{\mu}[\mathscr{A}] \Theta(x) )  .
\label{kin-equiv}
\end{align}
The equivalence (\ref{kin-equiv}) is shown from 
\begin{align}
 &  {\rm tr}(  ( {D}_{\mu}[\mathscr{A}] \Theta(x))^\dagger 	{D}^{\mu}[\mathscr{A}] \Theta(x) ) 
\nonumber\\
=& ({D}_{\mu}[\mathscr{A}]\tilde\Phi)^{\dagger}  \cdot ({D}^{\mu}[\mathscr{A}]\tilde\Phi) + ({D}_{\mu}[\mathscr{A}]\Phi)^{\dagger}  \cdot ({D}^{\mu}[\mathscr{A}]\Phi) 
 ,
\label{kin-equiv2}
\end{align}
by noting the equality, 
\begin{align}
 ({D}_{\mu}[\mathscr{A}]\Phi)^{\dagger}  \cdot ({D}^{\mu}[\mathscr{A}]\Phi) 
=  ({D}_{\mu}[\mathscr{A}]\tilde\Phi)^{\dagger}  \cdot ({D}^{\mu}[\mathscr{A}]\tilde\Phi) 
 ,
\label{kin-equiv3}
\end{align}
which follows from the fact that  $({D}_{\mu}[\mathscr{A}]\Phi)^{\dagger}  \cdot ({D}^{\mu}[\mathscr{A}]\Phi)$ is real-valued 
using (\ref{tilde-rel}) and 
$
\epsilon^t \epsilon=-\epsilon\epsilon=\bm{1}
$.


Thus the $SU(2)$ gauge-scalar model (\ref{SU2-gauge-scalar-f1})  is rewritten in terms of the matrix-valued scalar field as
\begin{align}
		\tilde{\mathscr{L}}_{\rm RF} =& -\frac{1}{2} {\rm tr}[ \mathscr{F} _{\mu\nu}(x)\mathscr{F}^{\mu\nu } (x)]
\nonumber\\& 
+ \frac{1}{2} {\rm tr}[  ( {D}_{\mu}[\mathscr{A}] \Theta(x))^\dagger 	{D}^{\mu}[\mathscr{A}] \Theta(x) ]
\nonumber\\& 
+ u(x) {\rm tr} \left(\Theta(x) ^{\dagger}\Theta(x) - \frac{1}{2}v^2 \bm{1} \right) /{\rm tr}(\bm{1}) .
\label{SU2-gauge-scalar-f2}
\end{align}
Notice that the $SU(2)$ gauge-scalar model (\ref{SU2-gauge-scalar-f2}) rewritten in terms of the matrix-valued scalar field $\Theta$ has the larger $SU(2)_{\rm local} \times SU(2)^\prime_{\rm global}$ symmetry than the original model,
\begin{align}
  \Theta(x) \to U(x) \Theta(x) U^\prime  , \ U \in SU(2)_{\rm local} , \ U^\prime \in  SU(2)^\prime_{\rm global} 
 .
 \label{SU2-custodial}
\end{align}
The extra global symmetry $SU(2)^\prime_{\rm global}$ is called the \textbf{custodial symmetry} which is a kind of flavor symmetry mixing the two scalar doublets. 
The custodial symmetry as a global symmetry can be broken spontaneously. The results coming from this fact will be discussed elsewhere.

\subsection{BEH mechanism 
for $SU(2)$ gauge-fundamental scalar model 
}

For the gauge group $SU(2)$, we introduce the vector boson field $\mathscr{W}_\mu$ defined in terms of the normalized scalar field $\hat{\Theta}$ and the original gauge field $\mathscr{A}_\mu$ as
\footnote{
By using $\hat{\Theta}(x)  \hat{\Theta}(x)^\dagger=\bm{1}$, we find the other expressions for $\mathscr{W}_\mu$,
\begin{align}
 \mathscr{W}_\mu(x) 
=&  ig^{-1}  \partial_{\mu} \hat{\Theta}(x) \hat{\Theta}(x)^\dagger +  \mathscr{A}_{\mu}(x) 
\nonumber\\
=& -ig^{-1} \hat{\Theta}(x) \partial_{\mu} \hat{\Theta}(x)^\dagger  +  \mathscr{A}_{\mu}(x)  
.
\label{W2}
\end{align}
}
\begin{align}
	\mathscr{W}_\mu(x) 
 :=& ig^{-1} ( {D}_{\mu}[\mathscr{A}] \hat{\Theta}(x)) \hat{\Theta}(x)^\dagger 
\nonumber\\
=& -ig^{-1} \hat{\Theta}(x)	({D}_{\mu}[\mathscr{A}] \hat{\Theta}(x) )^\dagger
\nonumber\\
=& \frac{1}{2} ig^{-1} [ ( {D}_{\mu}[\mathscr{A}] \hat{\Theta}(x)) \hat{\Theta}(x)^\dagger -  \hat{\Theta}(x)	({D}_{\mu}[\mathscr{A}] \hat{\Theta}(x) )^\dagger ]
 .
\label{W1-SU2}
\end{align}
The equivalence of the first two expressions in (\ref{W1-SU2}) follows from the Leibniz rule for the covariant derivative,
$
({D}_{\mu}[\mathscr{A}]\hat{\Theta}(x))  \hat{\Theta}(x)^\dagger+\hat{\Theta}(x)  ({D}_{\mu}[\mathscr{A}]\hat{\Theta}(x)^\dagger)
={D}_{\mu}[\mathscr{A}](\hat{\Theta}(x)  \hat{\Theta}(x)^\dagger)
=\partial_{\mu} (\bm{1})
=0$ 
using $\hat{\Theta}(x)  \hat{\Theta}(x)^\dagger=\bm{1}$.
By construction,  $\mathscr{W}_\mu$ transforms according to the adjoint representation under the gauge transformation,
\begin{align}
 \mathscr{W}_\mu(x)   \to \mathscr{W}_\mu^\prime(x) = U(x) \mathscr{W}_{\mu}(x) U(x)^\dagger
 .
 \label{W-transf}
\end{align}

We find that \textbf{the kinetic term of the scalar field $\Phi$ or $\Theta$ is identical to the mass term of the vector boson field $\mathscr{W}^\mu$ with the mass $M_W$}:
\begin{align}
 & ({D}_{\mu}[\mathscr{A}]\Phi)^{\dagger} ({D}^{\mu}[\mathscr{A}]\Phi) 
=  \frac{1}{2} {\rm tr}(  ( {D}_{\mu}[\mathscr{A}] \Theta(x))^\dagger 	{D}^{\mu}[\mathscr{A}] \Theta(x) ) 
\nonumber\\
=&  M_W^2 {\rm tr}( \mathscr{W}_\mu(x) \mathscr{W}^\mu(x) )  , 
\ M_W := \frac{1}{2}gv .
\label{SU2-mass-term}
\end{align}
It is remarkable that the mass term (\ref{SU2-mass-term}) of the vector boson field  $\mathscr{W}^\mu$ is gauge invariant.
Indeed, the transformation property (\ref{W-transf}) of $\mathscr{W}_\mu$ reconfirms the gauge invariance of the mass term (\ref{SU2-mass-term}) of $\mathscr{W}_\mu$. 
Thus, $\mathscr{W}_\mu$ defined by (\ref{W1-SU2}) is identified with the massive mode.%

In order to see the relationship between the new description and the conventional explanation for the BEH mechanism, we take the unitary gauge, namely, we can use the freedom of $SU(2)$ rotations to write the expectation value in the form:
\begin{align}
	&	\Phi(x)  
\to  \langle \Phi(x) \rangle = \Phi_\infty := \frac{1}{\sqrt{2}}
		 \begin{pmatrix}
			0 \\ 
			v
	\end{pmatrix} 
\nonumber\\& 
\Longleftrightarrow  
		\Theta(x)   
\to  \langle \Theta(x) \rangle = \Theta_\infty := \frac{1}{\sqrt{2}}
		  \begin{pmatrix} 
		v &	0 \\ 
		0 &	v
	\end{pmatrix} 
= \frac{v}{\sqrt{2}} \bm{1} .
\label{SU2-VEV}
\end{align}
By this choice of the vacuum expectation value of the scalar field, the original gauge symmetry $SU(2)$ is completely broken  with no residual gauge symmetry, which is called the \textbf{complete SSB}: $G=SU(2) \to H=\{1 \}$.%
\footnote{
If the matrix scalar field has the vacuum expectation value (\ref{SU2-VEV}), both $SU(2)$ and $SU(2)^\prime$ are broken, but the diagonal subgroup $SU(2)_{\rm diag}$ remains unbroken, see (\ref{SU2-custodial}).  
The original gauge-scalar model has the global symmetry $SU(2) \times SU(2)^\prime$ to be  spontaneously broken to $SU(2)_{\rm diag}$. Notice that only  $SU(2)$ in $SU(2) \times SU(2)^\prime$ is gauged in this model. 
}
In the complete SSB, all the components of the gauge boson become massive. 
This case should be compared with the \textbf{partial SSB}:  $G=SU(2) \to H=U(1)$ which occurs in the gauge-scalar model with an adjoint scalar field, as discussed in the previous paper \cite{Kondo16}.

In terms of the original scalar field, the kinetic term in the unitary gauge reduces to the mass term as  
\begin{align}  
 & ({D}_{\mu}[\mathscr{A}] \Phi)^\dagger  \cdot {D}^{\mu}[\mathscr{A}] \Phi
\nonumber\\  
\to&    [  ig  \Phi_\infty^\dagger \mathscr{A}_{\mu}  ]  \cdot
 [  -ig \mathscr{A}^{\mu}  \Phi_\infty ]  
\nonumber\\
=& g^2 \frac{v^2}{2} 
    \begin{pmatrix} 
		 0 & 1  \\ 
 	\end{pmatrix} \mathscr{A}_{\mu}  \mathscr{A}^{\mu} 
    \begin{pmatrix} 
		 0 \\ 
		 1
	\end{pmatrix} 
 = g^2 \frac{v^2}{2} (\mathscr{A}_{\mu}  \mathscr{A}^{\mu})_{22} 
\nonumber\\
=&  \frac{(gv)^2}{2} (T_A T_B)_{22} \mathscr{A}_{\mu}^A  \mathscr{A}^{\mu B} 
\nonumber\\
=&  \frac{(gv)^2}{4} (\{T_A ,T_B \} + [T_A ,T_B])_{22} \mathscr{A}_{\mu}^A  \mathscr{A}^{\mu ^B} 
\nonumber\\
=& \frac{1}{2}  \frac{(gv)^2}{4} \mathscr{A}_{\mu}^A  \mathscr{A}^{\mu ^A} 
 ,
\end{align} 
where we have used $\{T_A ,T_B \}=\frac{1}{2}\delta_{AB}$ for $T_A=\frac12 \sigma_A$ and $[T_A ,T_B]=-[T_B ,T_A]$.

In terms of the matrix-valued scalar field, similarly, the kinetic term in the unitary gauge reduces to the mass term  
\begin{align}
 & \frac{1}{2} {\rm tr}(  ( {D}_{\mu}[\mathscr{A}] \Theta(x))^\dagger 	{D}^{\mu}[\mathscr{A}] \Theta(x) ) 
  \nonumber\\
\to&  \frac{1}{2} {\rm tr}(    ig  {\Theta}_\infty^\dagger \mathscr{A}_{\mu}(x) 
 [  -ig \mathscr{A}^{\mu}(x)  {\Theta}_\infty ] ) 
\nonumber\\
=&  \frac{1}{2} g^2 \frac{v^2}{2} {\rm tr}(  \mathscr{A}_{\mu}(x) 
   \mathscr{A}^{\mu}(x)   ) .
\end{align}
In the unitary gauge, indeed, $\mathscr{W}_\mu$ reduces to the original gauge field,
\begin{align}
 \mathscr{W}_\mu(x) \to  ig^{-1} ( {D}_{\mu}[\mathscr{A}(x)] \hat{\Theta}_\infty) \hat{\Theta}_\infty^\dagger = \mathscr{A}_\mu(x) 
 .
\end{align}

We could have defined another gauge boson field $\tilde{\mathscr{W}}_\mu$ by
\begin{align}
 \tilde{\mathscr{W}}_\mu(x) &:= ig^{-1} \hat{\Theta}(x)^\dagger	{D}_{\mu}[\mathscr{A}] \hat{\Theta}(x) 
\nonumber\\
=& -ig^{-1} ( {D}_{\mu}[\mathscr{A}] \hat{\Theta}(x))^\dagger \hat{\Theta}(x) 
\nonumber\\
=& \frac{1}{2} ig^{-1} [ \hat{\Theta}(x)^\dagger	{D}_{\mu}[\mathscr{A}] \hat{\Theta}(x) -  ( {D}_{\mu}[\mathscr{A}] \hat{\Theta}(x))^\dagger \hat{\Theta}(x) ]
 .
 \label{W-tilde}
\end{align}
Notice that $\tilde{\mathscr{W}}_\mu$ and ${\mathscr{W}}_\mu$ are related as
\begin{align}
 \tilde{\mathscr{W}}_\mu(x)  =  \hat{\Theta}(x)^\dagger	   {\mathscr{W}}_\mu(x) \hat{\Theta}(x)   
 ,
 \label{TWT}
\end{align}
and that the vector boson field $\tilde{\mathscr{W}}_\mu$ is gauge invariant:
\begin{align}
 \tilde{\mathscr{W}}_\mu(x) \to \tilde{\mathscr{W}}_\mu^\prime(x) = \tilde{\mathscr{W}}_\mu(x) 
 .
\end{align}
The kinetic term of the scalar field $\Theta$ is equivalently rewritten into the mass term of the vector boson field $\tilde{\mathscr{W}}_\mu$:
\begin{align}
\frac{1}{2}  {\rm tr}(  ( {D}_{\mu}[\mathscr{A}] \Theta(x))^\dagger 	{D}^{\mu}[\mathscr{A}] \Theta(x) )
= M_W^2 {\rm tr}( \tilde{\mathscr{W}}_\mu(x) \tilde{\mathscr{W}}^\mu(x) ) ,
\end{align}
since 
${\rm tr}(  {\mathscr{W}}_\mu(x)  {\mathscr{W}}^\mu(x) )={\rm tr}( \tilde{\mathscr{W}}_\mu(x) \tilde{\mathscr{W}}^\mu(x) )$ from (\ref{TWT}).
However, the residual field defined by $\tilde{\mathscr{V}}_\mu:=\mathscr{A}_\mu-\tilde{\mathscr{W}}_\mu$ does not transform in a simple way for this choice of $\tilde{\mathscr{W}}_\mu$. 
This distinction does not occur for $U(1)$ gauge group.

\subsection{Field decomposition 
for $SU(2)$ gauge-fundamental scalar model
}

Once we identify $\mathscr{W}_\mu$ with the massive mode of the gauge field $\mathscr{A}_\mu$, the original gauge field $\mathscr{A}_\mu$ is separated into the massive vector field $\mathscr{W}_\mu$ and the residual one $\mathscr{V}_\mu$:
\begin{align}
	\mathscr{A}_\mu(x)  = \mathscr{V}_\mu(x) + \mathscr{W}_\mu(x) 
.
\label{A-decomposition}
\end{align}
Under the gauge transformation $U(x) \in SU(2)$, the original fields $\mathscr{A}_{\mu}$ and $\Theta$ transform as
\begin{align}
 & \mathscr{A}_{\mu}(x) \to U(x) \mathscr{A}_{\mu}(x) U(x)^\dagger + ig^{-1} U(x) \partial_\mu U(x)^\dagger  
 ,
\nonumber\\
 & \Theta(x) \to U(x) \Theta(x) .
\label{gauge-transf}
\end{align}
We have constructed  $\mathscr{W}_\mu$ so that it transform according to the adjoint representation,
\begin{align}
 \mathscr{W}_\mu(x)   \to U(x) \mathscr{W}_{\mu}(x) U(x)^\dagger
 .
\end{align}
Therefore, $\mathscr{V}_\mu$ transform just like the original gauge field,
\begin{align}
 & \mathscr{V}_{\mu}(x) \to U(x) \mathscr{V}_{\mu}(x) U(x)^\dagger + ig^{-1} U(x) \partial_\mu U(x)^\dagger  
.
\end{align}
To obtain the explicit expression for $\mathscr{V}_\mu$, we observe  that $\mathscr{W}_\mu=0$ yields the following condition for $\mathscr{V}_\mu$ up to the local gauge transformation:
\begin{align}
	 {D}_{\mu}[\mathscr{V}] \hat{\Theta}(x) =  \bm{0} \Leftrightarrow 
	 \partial_{\mu} \hat{\Theta}(x)   -ig \mathscr{V}_{\mu}(x) \hat{\Theta}(x)  =  \bm{0} 
 .
 \label{def-eq-SU2-1}
\end{align}
The \textbf{residual field} $\mathscr{V}_\mu$ is obtained by solving this equation    using $\hat{\Theta}  \hat{\Theta}^\dagger = \bm{1}$ as
\begin{align}
\mathscr{V}_{\mu}(x) =& -ig^{-1}\partial_{\mu} \hat{\Theta}(x) \hat{\Theta}(x)^\dagger  =  ig^{-1} \hat{\Theta}(x) \partial_{\mu}  \hat{\Theta}(x)^\dagger 
\nonumber\\ 
=& \frac{1}{2} ig^{-1} [ - \partial_{\mu}\hat{\Theta}(x)  \hat{\Theta}(x)^\dagger  
+  \hat{\Theta}(x)   \partial_{\mu} \hat{\Theta} ^\dagger   ]  ,
\nonumber\\ &
\  \hat{\Theta}(x) \in SU(2) .
\label{V-SU2-1}
\end{align}
This agrees with the result of (\ref{W2}). 
Of course, the residual field must be equal to 
$\mathscr{V}_\mu(x)=\mathscr{A}_\mu(x)-\mathscr{W}_\mu(x)$.
\footnote{
The residual field is written in terms of the doublet scalar field as
\begin{align}
\mathscr{V}_{\mu}(x) =&  ig^{-1} (\partial_\mu \hat\Phi(x) \hat\Phi^\dagger(x) + \epsilon \hat\Phi^*(x) \partial_\mu \hat\Phi^T(x) \epsilon^T ) , 
\nonumber\\ 
 \hat\Phi(x) :=&  \Phi(x)/\left(\frac{v}{\sqrt{2}}\right)   
 .
\label{V1b}
\end{align}
}

In the perturbative treatment the residual mode $\mathscr{V}_\mu$ is trivial. But it is non-trivial in the non-perturbative treatment, as discussed later.

\subsection{Reduction condition 
for $SU(2)$ Yang-Mills theory
}

In the $SU(2)$ gauge-scalar model,  $\mathscr{A}_\mu(x)$ and   $\Phi(x)$ are independent field variables. 
However, the pure $SU(2)$ Yang-Mills theory should be described by  $\mathscr{A}_\mu(x)$ alone and hence  $\Phi(x)$ must be supplied by the gauge field $\mathscr{A}_\mu(x)$  due to the strong interactions. 
In other words, the scalar field $\Phi(x)$ should be given as a (complicated) functional of the gauge field $\mathscr{A}_\mu(x)$.
This is achieved by imposing the appropriate constraint which we call the  \textbf{reduction condition}.

To find the reduction condition, we consider the extended gauge theory with the enlarged gauge symmetry $SU(2)_\omega \times SU(2)_\theta$ according to  the procedure given in \cite{KMS06}. 
The infinitesimal form of the enlarged gauge transformation is given by
\begin{align}
\delta_{\theta,\omega} \Theta(x) =& ig \theta(x) \Theta(x) , \ \theta(x) = \theta^A(x) T_A ,
\nonumber\\
\delta_{\theta,\omega} \mathscr{A}_{\mu}(x) =& \mathscr{D}_\mu[ \mathscr{A}] \omega(x) ,  \ \omega(x) = \omega^A(x) T_A .
\label{enlarge-g}
\end{align}
Under the enlarged gauge transformation (\ref{enlarge-g}), $\mathscr{W}_\mu$ transform as
\begin{align}
	\delta_{\theta,\omega} \mathscr{W}_\mu(x) 
 = \mathscr{D}_{\mu}[\mathscr{A}] \omega(x)  -   \mathscr{D}_{\mu}[\mathscr{V}] \theta(x)  
 .
\label{dW-SU2}
\end{align}
This is shown from   $\mathscr{W}_\mu=\mathscr{A}_\mu-\mathscr{V}_\mu$ by taking into account  (\ref{enlarge-g}) and 
\begin{align}
	\delta_{\theta,\omega} \mathscr{V}_\mu(x) 
=  \mathscr{D}_{\mu}[\mathscr{V}] \theta(x) 
 ,
\label{dW1-SU2b}
\end{align}
which is shown by applying  (\ref{enlarge-g}) to  (\ref{V-SU2-1}).
This is also shown by applying the enlarged gauge transformation to (\ref{W1-SU2}), although more  lengthy calculations are needed.
We can check that the enlarged gauge transformation recovers the infinitesimal form of the original gauge transformation when $\theta(x)=\omega(x)$:
\begin{align}
	\delta_{\omega} \mathscr{W}_\mu(x) 
 =& -ig [ \mathscr{W}_\mu(x) , \omega(x) ] 
 .
\label{dW2-SU2}
\end{align}
We obtain the reduction condition by minimizing a functional of the fields under the enlarged gauge transformation.
For our choice of the functional, the variation reads
\begin{align}
&	\delta_{\theta,\omega}  \int d^Dx \frac12 {\rm tr} \left( \mathscr{W}_\mu  \mathscr{W}_\mu \right) 
\nonumber\\ 
 =&	\int d^Dx {\rm tr} \left( \mathscr{W}_\mu \delta_{\theta,\omega} \mathscr{W}_\mu \right)
\nonumber\\ 
 =& \int d^Dx {\rm tr} \left( - \mathscr{W}_\mu  \mathscr{D}_{\mu}[\mathscr{V}] \theta  
+ \mathscr{W}_\mu \mathscr{D}_{\mu}[\mathscr{A}] \omega  \right)
\nonumber\\
 =& \int d^Dx  {\rm tr} \left( (\mathscr{D}_{\mu}[\mathscr{V}]  \mathscr{W}_\mu )  \theta - (\mathscr{D}_{\mu}[\mathscr{A}]  \mathscr{W}_\mu )  \omega   \right)
\nonumber\\ 
 =& \int d^Dx  {\rm tr} \left( (\theta - \omega ) (\mathscr{D}_{\mu}[\mathscr{A}]  \mathscr{W}_\mu )   \right)
\nonumber\\ 
 =& \int d^Dx  \frac{1}{2} (\theta - \omega )^A (\mathscr{D}_{\mu}[\mathscr{A}]  \mathscr{W}_\mu )^A  
  ,
\label{dW3-SU2}
\end{align}
where we have used the integration by parts in the third equality. 
For the functional to be minimized, 
 $(\mathscr{D}_{\mu}[\mathscr{A}]  \mathscr{W}_\mu )^A=0$ must be satisfied for $\theta \not=\omega$, while  
for $\theta =\omega$, this procedure imposes no condition.  By imposing the reduction condition, the enlarged gauge symmetry $SU(2)_\omega \times SU(2)_\theta$ is reduced to the original gauge symmetry $SU(2)_\alpha, \alpha=\theta=\omega$.
Therefore, the theory obtained by imposing the reduction condition has the $SU(2)$ local gauge symmetry. 
\footnote{
This procedure is regarded as the partial gauge fixing which breaks the enlarged gauge symmetry $SU(2)_\omega \times SU(2)_\theta$ into the original gauge symmetry $SU(2)$.
To be precise, this is the stationary condition. 
}
Thus we obtain the reduction condition:
\begin{align}
 \chi^A(x) :=& (\mathscr{D}^{\mu}[\mathscr{A}]  \mathscr{W}_\mu)^A(x) = 0 
\nonumber\\ 
   \Longleftrightarrow 
 \chi^A(x) =&  (\mathscr{D}^{\mu}[\mathscr{V}]  \mathscr{W}_\mu)^A(x) = 0 \ (A=1,2,3) 
 .
\label{SU2-reduction1}
\end{align}
\textbf{Imposing the reduction condition $\chi^A(x)=0$ ($A=1,2,3$)  eliminates three extra degrees of freedom introduced through a single radially fixed scalar field $\hat\Phi \in SU(2)$ (${\rm dim}SU(2)=3$), which is necessary to convert the $SU(2)$ gauge-scalar theory to the pure $SU(2)$ Yang-Mills theory.} 
\footnote{
Notice that the reduction condition is an off-shell condition. 
Therefore, solving the reduction condition is different from solving the field equation for the St\"uckelberg field as done in the preceding works \cite{KG67,Cornwall82,DTT88,RRA04}.
This means that the solution of the reduction condition does not necessarily satisfy the field equation. 
}
The reduction condition $\chi(x)=\chi^A(x) T_A$ is rewritten in terms of the   scalar field $\hat{\Theta}$ and the original gauge field $\mathscr{A}_\mu$ as
\begin{align}
   \chi(x) :=& \mathscr{D}_{\mu}[\mathscr{A}]  [( {D}^{\mu}[\mathscr{A}] \hat{\Theta}(x)) \hat{\Theta}(x)^\dagger] =  \bm{0}  
\Longleftrightarrow 
\nonumber\\  
 \chi(x) :=& - \mathscr{D}_{\mu}[\mathscr{A}][\hat{\Theta}(x)	({D}^{\mu}[\mathscr{A}] \hat{\Theta}(x) )^\dagger] =  \bm{0} 
 .
\label{SU2-reduction2}
\end{align}
The reduction condition is the gauge covariant equation, 
\begin{align}
 \chi(x)   \to U(x) \chi(x) U(x)^\dagger
 .
\end{align}
This implies that the reduction condition retains the same form under the gauge transformation, namely,  it is form-invariant.

\subsection{Field equations to the reduction condition 
for $SU(2)$ gauge-fundamental scalar model
}

We discuss the relationship between the reduction condition in the Yang-Mills theory and the field equation of the gauge-scalar model  described by  
\begin{align}
 \tilde{\mathscr{L}}_{\rm RF} =& -\frac{1}{2} {\rm tr}( \mathscr{F} _{\mu\nu}[\mathscr{A}]\mathscr{F}^{\mu\nu }[\mathscr{A}] )
\nonumber\\ &
 +  \frac{1}{2}{\rm tr}(  ( {D}_{\mu}[\mathscr{A}] \Theta )^\dagger 	{D}^{\mu}[\mathscr{A}] \Theta )  
\nonumber\\ &
+ u {\rm tr} \left(\Theta ^{\dagger}\Theta  - \frac{1}{2}v^2 \bm{1} \right) /{\rm tr}(\bm{1})  
 . 
\end{align}

For the $SU(2)$ gauge-scalar model with a radially fixed fundamental scalar field, the field equations are obtained by variation as
\footnote{
Notice that we have used the notation:  
$({D}^{\mu}[\mathscr{A}] \Theta)^\dagger \overleftarrow{{D}_{\mu}[\mathscr{A}]}^\dagger
=\{ \partial_{\mu} ( {D}^{\mu}[\mathscr{A}] \Theta )^\dagger + ( {D}^{\mu}[\mathscr{A}] \Theta )^\dagger  ig\mathscr{A}_{\mu} \}
$.
}
\begin{align}
0=& \frac{\delta \tilde{S}_{\rm RF}}{\delta u(x)} 
 = {\rm tr} \left(\Theta(x)^{\dagger}\Theta(x)  - \frac{1}{2}v^2 \bm{1} \right) /{\rm tr}(\bm{1})   ,
 \label{eq-u2}
\\ 
0=& \frac{\delta \tilde{S}_{\rm RF}}{\delta \Theta^\dagger(x)} 
 =  -  {D}_{\mu}[\mathscr{A}] {D}^{\mu}[\mathscr{A}] \Theta(x) + \Theta(x) u(x)   ,
 \label{eq-f2}
\\ 
0=& \frac{\delta \tilde{S}_{\rm RF}}{\delta \Theta (x)} 
 =  -   ({D}^{\mu}[\mathscr{A}] \Theta(x))^\dagger \overleftarrow{{D}_{\mu}[\mathscr{A}]}^\dagger +  u(x) \Theta(x)^\dagger   
 ,
 \label{eq-ff2}
\\ 
0=& \frac{\delta \tilde{S}_{\rm RF}}{\delta \mathscr{A}^\mu(x)} 
 =   \mathscr{D}^\nu [\mathscr{A}] \mathscr{F}_{\nu\mu}[\mathscr{A}](x) 
\nonumber\\ &
+ \frac{1}{2} ig [({D}_{\mu}[\mathscr{A}] \Theta(x)) {\Theta}(x)^\dagger  
-  {\Theta}(x) ({D}_{\mu}[\mathscr{A}] \Theta(x))^\dagger ]   ,
 \label{eq-A2}
\end{align}
where the field equation (\ref{eq-A2}) for $\mathscr{A}_\mu$ is equivalent to 
\begin{align}
0=& \frac{\delta \tilde{S}_{\rm RF}}{\delta \mathscr{A}^\mu(x)} 
 =   \mathscr{D}^\nu [\mathscr{A}] \mathscr{F}_{\nu\mu}[\mathscr{A}](x) +   M_W^2 \mathscr{W}_\mu(x)  
  .
 \label{eq-AAa2}
\end{align}

We proceed to study the relationship between the reduction condition and the field equation. 
Due to (\ref{eq-u2}), the scalar field $\Theta$ is normalized to obtain $\hat{\Theta}$. 
Multiplying (\ref{eq-f2}) by $\hat{\Theta}^\dagger$   and (\ref{eq-ff2}) by $\hat{\Theta}$ yields 
\begin{align}
0 =& \{ -{D}_{\mu}[\mathscr{A}]  ( {D}^{\mu}[\mathscr{A}] \hat{\Theta}) + \hat{\Theta} u \} \hat{\Theta}^\dagger
\nonumber\\ &
  -  \hat{\Theta} \{ - ({D}^{\mu}[\mathscr{A}] \hat{\Theta} )^\dagger \overleftarrow{{D}_{\mu}[\mathscr{A}]}^\dagger + u \hat{\Theta}^\dagger \} 
=  2ig \mathscr{D}_\mu[\mathscr{A}]  \mathscr{W}^\mu  
 \label{eq-AA-SU2}
.
\end{align}
Applying the covariant derivative to (\ref{eq-A2}) or (\ref{eq-AAa2}) yields 
\begin{align}
 0 = \mathscr{D}_\mu[\mathscr{A}] (\mathscr{D}_\nu [\mathscr{A}] \mathscr{F}^{\nu\mu}[\mathscr{A}] + M_W^2  \mathscr{W}^\mu  ) = M_W^2 \mathscr{D}_\mu[\mathscr{A}]    \mathscr{W}^\mu 
 \label{eq-AA3-SU2}
.
\end{align}
Thus we can draw an important conclusion: 
\textbf{If the fields $\mathscr{A}$ and $\Theta$ are a set of solutions of the field equations for the $SU(2)$ gauge-scalar model with a radially fixed fundamental scalar field, they automatically satisfy the reduction condition (\ref{SU2-reduction1}) for the pure $SU(2)$ Yang-Mills theory (with the gauge-invariant mass term)}.
Incidentally, the vector $\mathscr{W}^\mu$ in the non-Abelian case is not proportional to the Noether current $J^\mu$ associated to the global symmetry $SU(N)$.

We consider which field configuration can be a solution of the field equations. 
If the field $\mathscr{A}(x)$ is the solution of the self-dual equation $\mathscr{F}_{\mu\nu}[\mathscr{A}](x)=\pm {}^*\mathscr{F}_{\mu\nu}[\mathscr{A}](x)$, then it is automatically a solution of the Yang-Mills field equation 
$\mathscr{D}^\nu [\mathscr{A}] \mathscr{F}_{\nu\mu}[\mathscr{A}](x)=0$ due to the Bianchi identity $\mathscr{D}^\nu  [\mathscr{A}] {}^*\mathscr{F}_{\nu\mu}[\mathscr{A}](x)=0$. 
If the field $\mathscr{A}(x)$ was a configuration satisfying the self-dual condition, the massive vector boson must vanish identically, $\mathscr{W}_\mu(x) \equiv 0$, to satisfy the field equation (\ref{eq-AAa2}): 
\begin{align}
 & \mathscr{F}_{\mu\nu}[\mathscr{A}](x)=\pm {}^*\mathscr{F}_{\mu\nu}[\mathscr{A}](x)
 \Longrightarrow \mathscr{D}^\nu [\mathscr{A}] \mathscr{F}_{\nu\mu}[\mathscr{A}](x)=0
\nonumber\\ & 
\Longrightarrow \mathscr{W}_\mu(x) \equiv 0 
  .
\end{align}
Therefore, the instanton in the pure Yang-Mills theory cannot be a solution of the field equation (\ref{eq-AAa2}) of the gauge-scalar model. 
In the large (Euclidean) distance $\sqrt{x^2} \to \infty$, however, $\mathscr{W}_\mu$ falls off $\mathscr{W}_\mu(x) \to 0$ and  the field equation reduces to that of the ordinary massless Yang-Mills theory which is satisfied by  $\mathscr{V}_\mu(x)$,
\begin{align}
 & \mathscr{W}_\mu(x) \to 0
 \Longrightarrow \mathscr{D}^\nu [\mathscr{V}] \mathscr{F}_{\nu\mu}[\mathscr{V}](x) \to 0
  .
\end{align}
We suppose that the residual mode $\mathscr{V}_\mu(x)$ is given by the self-dual configuration or instanton (and antiinstanton) on whole spacetime $\mathscr{F}_{\mu\nu}[\mathscr{V}](x)=\pm {}^*\mathscr{F}_{\mu\nu}[\mathscr{V}](x)$ and that the discrepancy can be cared by the massive mode $\mathscr{W}_\mu(x)$ in such a way that the sum $\mathscr{V}_\mu(x)+\mathscr{W}_\mu(x)$ reproduces the solution for $\mathscr{A}_\mu(x)$. 
This strategy greatly facilitates finding the solution of the gauge-scalar model.  This is an advantage of decomposing the original gauge field $\mathscr{A}_\mu(x)$ into the two pieces $\mathscr{V}_\mu(x)$ and $\mathscr{W}_\mu(x)$.  
The explicit solution based on this observation will be given in a subsequent paper.

\subsection{Representations in terms of original scalar fields}

In order to obtain the expressions in terms of the original  scalar field $\Phi$, it is sufficient to impose the condition:
\begin{align}
	 {D}_{\mu}[\mathscr{V}] \hat{\Phi}(x) =  0 . 
\label{red0}
\end{align}
In fact, (\ref{def-eq-SU2-1}) follows from (\ref{red0}):
\begin{align}
& {D}_{\mu}[\mathscr{V}] \hat{\Phi}(x)  = 0 \Longrightarrow  {D}_{\mu}[\mathscr{V}] \hat{\Theta}(x) = 0
 ,
\label{def-V-step1}
\end{align}
since
\begin{align}
  {D}_{\mu}[\mathscr{V}] \hat{\Theta}  
 =&  \partial_\mu \hat{\Theta} -ig \mathscr{V}_\mu \hat{\Theta}
\nonumber\\
=& 	\partial_\mu 
	  \begin{pmatrix} 
		\hat{\tilde\Phi} &	\hat{\Phi} 
	\end{pmatrix} 
-ig \mathscr{V}_\mu 
	  \begin{pmatrix} 
		\hat{\tilde\Phi} &	\hat{\Phi} 
	\end{pmatrix} 
\nonumber\\
=& 	 
	  \begin{pmatrix} 
		{D}_{\mu}[\mathscr{V}] \hat{\tilde\Phi} & {D}_{\mu}[\mathscr{V}] \hat{\Phi} 
	\end{pmatrix} 
 ,
\end{align}
and
\begin{align}
 \epsilon ({D}_{\mu}[\mathscr{V}] \hat{\Phi})^*
 =&  \epsilon (\partial_\mu \hat{\Phi} -ig \mathscr{V}_\mu \hat{\Phi})^*
\nonumber\\
=& \partial_\mu (\epsilon \hat{\Phi}^*) -ig \epsilon \mathscr{V}_\mu^* \epsilon \epsilon \hat{\Phi}^* 
\nonumber\\
=& \partial_\mu (\epsilon \hat{\Phi}^*) -ig  \mathscr{V}_\mu (\epsilon \hat{\Phi}^*) 
= {D}_{\mu}[\mathscr{V}] \hat{\tilde\Phi}
 ,
\end{align}
where we have used 
$
\epsilon\epsilon=-\bm{1}
$ 
and 
$
\epsilon \sigma_A^* \epsilon = \sigma_A
$.

For $G=SU(2)$, the defining equation (\ref{red0}) for the residual field $\mathscr{V}$ is given by
\begin{align}
 \partial_{\mu} \hat{\Phi}(x)   -ig \mathscr{V}_{\mu}(x) \hat{\Phi}(x)   = 0 .
\label{red1}
\end{align}
Multiplying (\ref{red1}) by $\hat{\Phi}^\dagger T_A$ from the left yields 
\begin{align}
 	 \hat{\Phi}^\dagger T_A \partial_{\mu} \hat{\Phi}   -ig \mathscr{V}_{\mu}^B \hat{\Phi}^\dagger T_A T_B \hat{\Phi}    = 0 
 .
 \label{eq1a}
\end{align}

Taking the adjoint of (\ref{red1}), on the other hand, we obtain
\begin{align}
 \partial_{\mu} \hat{\Phi} ^\dagger   +ig  \hat{\Phi} ^\dagger \mathscr{V}_{\mu}   = 0 
 .
 \label{red1b}
\end{align}
Multiplying  (\ref{red1b}) by $T_A \hat{\Phi}$ from the right leads to 
\begin{align}
 \partial_{\mu} \hat{\Phi}^\dagger T_A \hat{\Phi}  +ig \mathscr{V}_{\mu}^B \hat{\Phi}^\dagger T_B T_A \hat{\Phi}   = 0 
 .
 \label{eq2a}
\end{align}
By subtracting (\ref{eq1a}) from (\ref{eq2a}), we obtain
\begin{align}
 & \hat{\Phi}^\dagger T_A \partial_{\mu} \hat{\Phi} - \partial_{\mu} \hat{\Phi}^\dagger T_A \hat{\Phi} 
 -ig \mathscr{V}_{\mu}^B \hat{\Phi}^\dagger \{ T_A , T_B \} \hat{\Phi}    
  = 0 
 ,
\end{align}
which is rewritten into 
\begin{align}
 & \hat{\Phi}^\dagger T_A \partial_{\mu} \hat{\Phi} - \partial_{\mu} \hat{\Phi}^\dagger T_A \hat{\Phi} 
 -ig \frac12 \mathscr{V}_{\mu}^A  
   = 0 
 ,
\end{align}
where we have used the relation 
$
 \{ T_A , T_B \} := T_A T_B + T_B T_A  =  \frac12 \delta_{AB} \bm{1}  
$
for the $SU(2)$ generators $T_A=\frac{\sigma^A}{2}$, and $\hat{\Phi}^\dagger \hat{\Phi}=1$.
Thus we obtain
\begin{align}
   g \frac12 \mathscr{V}_{\mu}^A(x)   
= -i [\hat{\Phi}^\dagger(x) T_A \partial_{\mu} \hat{\Phi}(x) - \partial_{\mu} \hat{\Phi}^\dagger(x) T_A \hat{\Phi}(x) ]
  ,
\nonumber
\end{align}
or
\begin{align}
   g \mathscr{V}_{\mu}^A(x)  
= -i [\hat{\Phi}^\dagger(x) \sigma_A \partial_{\mu} \hat{\Phi}(x) - \partial_{\mu} \hat{\Phi}^\dagger(x) \sigma_A \hat{\Phi}(x) ]
  .
  \label{V-1b}
\end{align}
This expression for the residual field agrees with the previous one (\ref{V-SU2-1}).

For $G=SU(2)$, thus, the gauge field is decomposed as
\begin{align}
 \mathscr{A}_{\mu}^A(x) =&   \mathscr{W}_{\mu}^A(x) +  \mathscr{V}_{\mu}^A(x)  ,
\nonumber\\ 
 \mathscr{V}_{\mu}^A(x)  
=& -ig^{-1} [\hat{\Phi}^\dagger(x) \sigma_A \partial_{\mu} \hat{\Phi}(x) - \partial_{\mu} \hat{\Phi}^\dagger(x) \sigma_A \hat{\Phi}(x) ]
  ,
\nonumber\\
 \mathscr{W}_{\mu}^A(x)  
=& ig^{-1} [\hat{\Phi}^\dagger(x) \sigma_A D_{\mu}[\mathscr{A}] \hat{\Phi}(x) 
\nonumber\\&
  - (D_{\mu}[\mathscr{A}] \hat{\Phi}(x))^\dagger \sigma_A \hat{\Phi}(x) ]
  .
  \label{W-A1}
\end{align}
In fact, summing up $\mathscr{W}_{\mu}$ and $\mathscr{V}_{\mu}$ recovers the original gauge field $\mathscr{A}_{\mu}$:
\begin{align}
  \mathscr{V}_{\mu}^A + \mathscr{W}_{\mu}^A  
=& ig^{-1} [\hat{\Phi}^\dagger \sigma_A  (-ig\mathscr{A}_\mu \hat{\Phi}) 
 - (-ig\mathscr{A}_\mu \hat{\Phi})^\dagger \sigma_A \hat{\Phi} ]
\nonumber\\
=&  \hat{\Phi}^\dagger \sigma_A  \mathscr{A}_\mu  \hat{\Phi} + \hat{\Phi}^\dagger \mathscr{A}_\mu  \sigma_A \hat{\Phi} 
\nonumber\\
=&  \hat{\Phi}^\dagger \frac{1}{2} \{ \sigma_A , \sigma_B \}  \hat{\Phi}   \mathscr{A}_\mu^B
\nonumber\\
=&  \hat{\Phi}^\dagger  \hat{\Phi}   \mathscr{A}_\mu^A
= \mathscr{A}_\mu^A 
  ,
  \label{A-A1}
\end{align}
where we have used
$
 \{ \sigma_A , \sigma_B \} = 2 \delta_{AB} \bm{1}
$.
We find that in the limit of taking the uniform (or constant) scalar field, $\mathscr{V}_{\mu}$ vanishes and $\mathscr{W}_{\mu}$ approaches $\mathscr{A}_{\mu}$:
\begin{align}
\Phi(x) \to \Phi 
 \Longrightarrow 
 \mathscr{V}_{\mu}^A(x)  \to 0, 
\
 \mathscr{W}_{\mu}^A(x)  \to \mathscr{A}_{\mu}^A(x) .
\end{align}

\subsection{Change of variables and reformulation of Yang-Mills theory}

The partition function of the gauge-scalar model with the radially fixed constraint, \begin{align}
f(\Phi(x)):=\Phi(x) ^{\dagger} \Phi(x)  - \frac{1}{2}v^2=0,
\end{align} 
is defined by 
\begin{align}
 Z_{\rm RF} =& \int \mathcal{D} \mathscr{A}  \mathcal{D} \Phi \mathcal{D}u  e^{iS_{\rm RF}[\mathscr{A},\Phi,u]}
 \nonumber\\
   =& \int \mathcal{D} \mathscr{A}  \mathcal{D}\Phi  \prod_{x} \delta\left( f(\Phi(x))  \right) e^{iS_{\rm RF}[\mathscr{A},\Phi]} 
 \nonumber\\
   =& \int \mathcal{D} \mathscr{A}  \mathcal{D}\hat\Phi e^{iS_{\rm RF}[\mathscr{A},\Phi]} ,
\label{Z-massive-YM1}
\end{align}
where the action is given by
\begin{align}
S_{\rm RF}[\mathscr{A},\Phi]=S_{\rm YM}[\mathscr{A}]+S_{\rm kin}[\mathscr{A},\Phi] ,
\end{align}
and the integration measures are given by
\begin{align}
 & \mathcal{D}\mathscr{A}:=\prod_{x,\mu,A}[d\mathscr{A}_\mu^A(x)], \
\mathcal{D}\Phi :=\prod_{x}[d\Phi(x)],  
 \nonumber\\ &
\mathcal{D}u :=\prod_{x}[du(x)] .
\end{align}

In order to obtain the Yang-Mills theory with a gauge-invariant mass term by starting from the corresponding ``complementary'' gauge-scalar model, we must eliminate the extra degrees of freedom which are brought into the Yang-Mills theory by the (radially fixed) scalar field. 
For this purpose, we restrict the field configuration space $(\mathscr{A},\Phi)$ to the subspace subject to the appropriate constraint $\bm{\chi}=0$ which we call the \textbf{reduction condition}.
Here the reduction condition $\chi=0$ is understood to be written in terms of $\mathscr{A}$ and $\Phi$, $\chi=\chi[\mathscr{A},\Phi]$, see (\ref{SU2-reduction2}). 
Following the way similar to the Faddeev-Popov procedure, we insert the unity to the functional integral:
\begin{align}
  1 = \int \mathcal{D} \bm{\chi}^\theta \delta(\bm{\chi}^\theta)
=   \int\! \mathcal{D}\bm\theta \delta(\bm\chi^\theta) \Delta^{\rm red} ,
\end{align}
where 
$\bm{\chi}^\theta:=\bm{\chi}[ \mathscr{A},\Phi^\theta]$ 
is the reduction condition written in terms of $\mathscr{A}$ and  $\Phi^\theta$ which is the local rotation of $\Phi$ by $\bm\theta$)
and 
$
\Delta^{\rm red}:=\det\left(\frac{\delta\bm\chi^\theta}{\delta{\bm\theta}}\right)
$
denotes  the Faddeev-Popov determinant associated with the reduction condition  $\chi=0$, see  \cite{KKSS15} for the details. 
Note that $\bm\theta$ have the same degrees of freedom as $\bm\chi$. 
Then we obtain
\begin{align}
 Z_{\rm RF} =&  \int \mathcal{D}\hat\Phi   \mathcal{D} \mathscr{A} \int\! \mathcal{D}\bm\theta   \delta(\bm\chi^\theta) \Delta^{\rm red}   e^{iS_{\rm RF}[\mathscr{A},\Phi ]} .
\label{Z-gauge-scalar-f1}
\end{align}

We perform the change of variables from the original variables $(\hat\Phi^a, \mathscr{A}_\mu^A )$ to the new variables 
$(\hat\Phi^b, \mathscr{W}_\nu^B )$:
\begin{align}
 (\hat\Phi^a, \mathscr{A}_\mu^A ) \to (\hat\Phi^b,\mathscr{W}_\nu^B) .
\end{align}
Then the partition function reads
\begin{align}
 Z_{\rm mYM} =& \int \mathcal{D} \hat\Phi  \mathcal{D}\mathscr{W} {J} \int \mathcal{D}\bm\theta   \delta(\tilde{\bm\chi}^\theta) \tilde\Delta^{\rm red}   e^{i\tilde{S}_{\rm mYM}[\mathscr{W},\hat\Phi ]} ,
\label{Z}
\end{align}
where the action $\tilde{S}_{\rm mYM}[\mathscr{W},\hat\Phi]$ with the gauge-invariant mass term $S_{\rm m}[\mathscr{W}]$ is obtained by substituting the decomposition of $\mathscr A_\mu$ into $S_{\rm RF}[\mathscr{A},\Phi ]$:
\begin{align}
\tilde{S}_{\rm mYM}[\mathscr{W},\hat\Phi]=S_{\rm YM}[\mathscr{V}+\mathscr{W}]+S_{\rm m}[\mathscr{W}] ,
\end{align}
and the integration measure is given by
\begin{align}
\mathcal{D}\mathscr{W}:=\prod_{x,\mu,A}[d\mathscr{W}_\mu^A(x)] .
\end{align}
with  the Jacobian ${J}$ associated with change of variables from 
$(\hat\Phi^a, \mathscr{A}_\mu^A )$ to  
$(\hat\Phi^b, \mathscr{W}_\nu^B )$,
\begin{align}
\mathcal{D} \hat\Phi  \mathcal{D} \mathscr{A}   
=  {J} \mathcal{D}\hat\Phi  \mathcal{D}\mathscr{W}  , 
\quad 
 {J} =  
 \begin{vmatrix}
 \frac{\partial \Phi^a}{\partial \hat\Phi^b} &  \frac{\partial \Phi^a}{\partial \mathscr{W}_\nu^B}   \cr
   \frac{\partial \mathscr{A}_\mu^A}{\partial \hat\Phi^b} & \frac{\partial \mathscr{A}_\mu^A}{\partial \mathscr{W}_\nu^B} \cr
 \end{vmatrix} 
   .
\end{align}
Here the reduction condition $\tilde\chi=0$ and the associated determinant $\tilde\Delta^{\rm red}$ are  supposed to be written in terms of $\mathscr{W}$    and $\hat\Phi$, see (\ref{SU2-reduction1}).

Moreover, we perform the change of variables $\Phi \rightarrow \Phi^{\theta}$, i.e., the local rotation by the angle $\theta$ and the corresponding gauge transformation    for the other new variables $\mathscr{W}_\mu$: 
$\mathscr{W}_\mu \rightarrow \mathscr{W}_\mu^{\theta}$. 
From the gauge invariance of the action $\tilde S_{\rm mYM}[\mathscr{W},\hat\Phi]$ and the integration measure 
$\mathcal{D}\hat\Phi\mathcal{D}\mathscr{W}_\mu$, 
we can rename the dummy integration variables $\Phi^{\theta} , \mathscr{W}_\mu^{\theta}$  as $\Phi, \mathscr{W}_\mu$ respectively.
Thus the integrand does not depend on $\theta$ and the gauge volume $\int\! \mathcal{D}\bm\theta$ can be factored out:%
\begin{align}
  {Z}_{{\rm mYM}}  
=& \int\! \mathcal{D}\bm\theta 
\int \mathcal{D}\hat\Phi \mathcal{D}\mathscr{W}   
 {J}
\delta(\hat{\bm\chi}) \tilde\Delta ^{\rm red} 
 e^{i\tilde{S}_{\rm mYM}[\mathscr{W},\hat\Phi]} .
\end{align}
Note that the Faddeev--Popov determinant 
 $\tilde\Delta ^{\rm red}$ can be rewritten into another form:  
\begin{equation}
 \tilde\Delta ^{\rm red} 
:= \det\left(\frac{\delta \hat{\bm\chi}}{\delta{\bm\theta}}\right)_{\bm{\chi}=0}
=   \det\left(\frac{\delta \hat{\bm\chi}}{\delta \Phi^\theta}\right)_{\bm{\chi}=0}.
\end{equation}
Ignoring the gauge volume $\int\! \mathcal{D}\bm\theta$, thus, we have arrived at the reformulated Yang-Mills theory  in which the independent variables are regarded as $\hat\Phi(x)$ and $\mathscr W_\mu(x)$ with the partition function:
\begin{align}
 {Z}_{{\rm mYM}}^\prime  
=&   \int \mathcal{D}\hat\Phi \mathcal{D}\mathscr{W}        
  {J} 
\delta(\tilde{\bm\chi})  
   \tilde\Delta ^{\rm red}
 e^{i\tilde{S}_{\rm mYM}[\mathscr{W},\hat\Phi ]} , 
\end{align}
where the constraint is rewritten in terms of the new variables:
\begin{equation}
\tilde{\bm\chi} 
 =\tilde{\bm\chi} [\mathscr{W},\hat\Phi]
 :=\mathscr{D}^\mu[\mathscr{V} ]\mathscr{W}_\mu .
\end{equation}

Now we show that the Jacobian $J$ is a field-independent numerical factor.
Since $\mathscr{W}_\nu^B$ and $\hat\Phi^b $ are independent, we have
\begin{align}
  \frac{\partial \hat\Phi^a}{\partial \hat\Phi^b} = \delta^{ab}  , \quad
  \frac{\partial \hat\Phi^a}{\partial \mathscr{W}_\nu^B} = 0 .
\end{align}
Then the Jacobian is reduced to the determinant of the $3D\times 3D$ matrix:
\begin{align}
 {J} = 
 \begin{vmatrix}
  \delta^{ab} & 0    \cr
  \frac{\partial \mathscr{A}_\mu^A}{\partial \hat\Phi^b} &  \frac{\partial \mathscr{A}_\mu^A}{\partial \mathscr{W}_\nu^B}  \cr
 \end{vmatrix}  
=  \begin{vmatrix}
   \frac{\partial \mathscr{A}_\mu^A}{\partial \mathscr{W}_\nu^B} \cr
 \end{vmatrix} .
\end{align}
This implies that $J$ is independent of how $\hat\Phi$ is related to the original field $\mathscr{A}$, that is to say, $J$ does not depend on the choice of the reduction condition, since the Jacobian does not depend on $\frac{\partial \mathscr{A}_\mu^A}{\partial \hat\Phi^b}$.  
In order to calculate $\frac{\partial \mathscr{A}_\mu^A}{\partial \mathscr{W}_\nu^b}$, 
we rewrite $\mathscr{A}_\mu^A$ in terms of independent degrees of freedom $(\hat\Phi^b,\mathscr{W}_\nu^B)$. 
The field $\mathscr{A}_\mu^A$ is decomposed into $\mathscr{W}_\mu^A$ and $\mathscr{V}_\mu^A$, i.e., 
$
\mathscr{A}_\mu^A =  \mathscr{V}_\mu^A + \mathscr{W}_\mu^A 
$, 
and $\mathscr{V}_\mu^A$ is written in terms of $\hat\Phi$ alone. 
Therefore, we have
\begin{align}
  \frac{\partial \mathscr{A}_\mu^A}{\partial \mathscr{W}_\nu^B} = \delta_{\mu\nu} \delta_B^A .
\end{align}
Thus, we conclude that the Jacobian reduced to a field-independent numerical factor:
\begin{align}
 {J} = 
  \begin{vmatrix}
    \delta_{\mu\nu} \delta_B^A  \cr
 \end{vmatrix}
=   
  \begin{vmatrix}
      \delta_B^A \cr
 \end{vmatrix}^D
= 1  .
\label{C26-J=1}
\end{align}

\subsection{Implications for quark confinement}

For confinement of colored objects to occur, there must exist the long-range confining force which could be mediated by massless gluons.  
In the  $SU(2)$ gauge-scalar model with AN adjoint scalar there remains a massless gauge field even after the two components among three components of the gauge field become massive by the BEH mechanism. Therefore, the residual mode  contains a massless gauge field. However, the residual gauge field contains also the color direction field $\bm{n}$ which describes magnetic monopole in the Yang-Mills theory. 
In the three-dimensional spacetime, it has been shown by an analytical way that quark confinement occurs due to magnetic monopoles in the Georgi-Glashow model \cite{Polyakov77b}, which can be extended \cite{NMWK18} to the pure Yang-Mills theory with a gauge-invariant gluon mass generated according to the gauge-independent BEH mechanism \cite{Kondo16}. 
In the four-dimensional spacetime, it has been confirmed by  numerical simulations that closed loops of magnetic monopole which are identified with monopole-antimonopole pairs are dominant configurations responsible for quark confinement, see e.g. \cite{KKSS15} and references therein.

In the $SU(2)$ gauge-scalar model with a fundamental scalar,  however, such massless gluons mediating the long-range force no longer exist after the BEH mechanism occurs, since all the components of gluons become massive. 
Therefore, the field mode responsible for the long-range confining force must be attributed to the residual mode $\mathscr{V}$. 
From this point of view, topological defects represented by the residual mode  of the pure gauge form can be the promising candidates for topological objects mediating long-range confining force.


\subsection{Topology 
for $SU(2)$ gauge-fundamental scalar model
}

Notice that $\mathscr{V}_{\mu}$ is of the pure gauge. 
In the topologically trivial sector, $\mathscr{V}_\mu$ vanishes in the unitary gauge (except the singular points),
$\mathscr{V}_{\mu}(x) \to  -ig^{-1}\partial_{\mu} \hat{\Theta}_\infty \hat{\Theta}^\dagger_\infty  = 0 
 .
$
Rather, this part can give the nonvanishing topological configurations such as  instantons and magnetic monopole current see, e.g., Ref. \cite {Kondo97}. 
The target space of the scalar field (vacuum manifold) is $SU(2) \simeq S^3$. Therefore, we consider the map $\Phi: S_\infty^n \to S^3$ from the $n$-dimensional sphere $S^n$ at infinity in the $D$-dimensional space-time to the vacuum manifold $SU(2)$.  Then the topological non-trivial configuration is characterized by the non-trivial homotopy group $\Pi_n(SU(2))=\Pi_n(S^3)\not= 0$. 
The non-trivial homotopy is possible only when $n \ge 3$, especially for $n=3$, $\Phi: S_\infty^3 \to S^3$, namely, the $SU(2)$ field defined on a three-sphere $S_\infty^3$ in the space-time with the non-trivial homotopy $\Pi_3(S^3) = \mathbb{Z}$. 
Notice that
$\Pi_1(S^3)=\Pi_2(S^3)=0$.

\section{Color direction field from the fundamental scalar field}

The color direction field $\bm{n}(x)$ plays the key role for defining the new field variables and giving  gauge-invariant magnetic monopoles in the reformulated Yang-Mills theory \cite{KKSS15}. 
In the adjoint scalar case, the color direction field agrees with the normalized scalar field, i.e., $\bm{n}(x)=\hat{\bm{\phi}}(x)$, which is reasonable because the color direction field transforms in the adjoint representation under the gauge transformation. 
In the fundamental scalar case, however, the color field must be constructed as a composite operator of the fundamental scalar field.

\subsection{Color direction field}

We introduce the unit vector field $\bm{n}(x)$ with three components $n_A(x)$ $(A=1,2,3)$ satisfying $n_A(x)n_A(x)=1$ which we call the \textbf{color direction field} or \textbf{color field} in short. 
In fact, the color direction field plays the key role for giving  gauge-invariant magnetic monopoles \cite{KKSS15}. 
We can construct the color direction field $n_A$ by using a doublet of the complex scalar field $\Phi$ as 
\begin{align}
  n_A(x) :=& \mp 2 \hat{\Phi}(x)^\dagger T_A \hat{\Phi}(x) 
= \mp \hat{\Phi}(x)^\dagger \sigma_A \hat{\Phi}(x) 
\nonumber\\
=& \mp  \hat\Phi^{\ast}_a(x) (\sigma_A)_{ab} \hat\Phi_b(x) , \ (a,b=1,2)
  ,
  \label{def-color-f1}
\end{align}
where $\sigma_A$ are Pauli matrices. 
 Each component reads 
\begin{align}
 n_1 =& \mp (\hat{\bm{\phi}}^{\ast}_1 \hat{\bm{\phi}}_2 + \hat{\bm{\phi}}_1\hat{\bm{\phi}}^{\ast}_2) 
=\mp 2\mathrm{Re}(\hat{\bm{\phi}}^{\ast}_1\hat{\bm{\phi}}_2) 
\nonumber\\
=& \pm 2(\hat{ {\phi}}_1\hat{ {\phi}}_3-\hat{ {\phi}}_2\hat{ {\phi}}_0)  ,
\nonumber\\
 n_2 =&  \mp i(-\hat{\bm{\phi}}^{\ast}_1\hat{\bm{\phi}}_2 + \hat{\bm{\phi}}_1\hat{\bm{\phi}}^{\ast}_2) 
= \mp 2\mathrm{Im}(\hat{\bm{\phi}}^{\ast}_1\hat{\bm{\phi}}_2) 
\nonumber\\
=&  \pm 2(\hat{ {\phi}}_2\hat{ {\phi}}_3+\hat{ {\phi}}_1\hat{ {\phi}}_0)   ,
\nonumber\\
 n_3 =& \mp (\hat{\bm{\phi}}^{\ast}_1\hat{\bm{\phi}}_1 - \hat{\bm{\phi}}^{\ast}_2\hat{\bm{\phi}}_2 ) = \mp (|\hat{\bm{\phi}}_1|^2-|\hat{\bm{\phi}}_2|^2 )  
\nonumber\\
=&  \pm ( -\hat{ {\phi}}^2_1-\hat{ {\phi}}^2_2+\hat{ {\phi}}^2_3+\hat{ {\phi}}^2_0) .
\end{align}
This color field is reproduced in terms of the matrix-valued scalar field $\Theta$:
\begin{align}
 \bm{n}(x)  
= \pm \hat\Theta(x) \sigma_3 \hat\Theta(x)^{\dagger} , \  \hat\Theta(x) \in SU(2) , 
  \label{def-color-f2}
\end{align}
if it is identified with the Lie-algebra valued field:
\begin{equation}
 \bm{n}(x) := n_A(x) \sigma_A =  
  \begin{pmatrix}
   n_3(x) & n_1(x)-in_2(x) \\
   n_1(x)+in_2(x) & -n_3(x) \\
  \end{pmatrix}
   .
\end{equation} 
The equivalence between (\ref{def-color-f1}) and (\ref{def-color-f2}) is shown by explicit calculations: for example, using (\ref{Theta1}) and (\ref{Theta2}), we have 
\begin{align}
 \bm{n}  
=& \pm \hat\Theta \sigma_3 \hat\Theta^{\dagger} 
  \nonumber\\
=&  \pm 
    \begin{pmatrix} 
	 \hat{\bm{\phi}}_2^* &	\hat{\bm{\phi}}_1 \\ 
     -\hat{\bm{\phi}}_1^* & \hat{\bm{\phi}}_2
	\end{pmatrix}
\begin{pmatrix}
 1 & 0 \\
 0 & -1 \\
 \end{pmatrix} 
	 \begin{pmatrix} 
		\hat{\bm{\phi}}_2 & -\hat{\bm{\phi}}_1 \\ 
		\hat{\bm{\phi}}_1^* &	\hat{\bm{\phi}}_2^*
	\end{pmatrix} 
  \nonumber\\
=&  \pm 
    \begin{pmatrix} 
		\hat{\bm{\phi}}_2^* &	\hat{\bm{\phi}}_1 \\ 
		-\hat{\bm{\phi}}_1^* & \hat{\bm{\phi}}_2
	\end{pmatrix}
	 \begin{pmatrix} 
		\hat{\bm{\phi}}_2 & -\hat{\bm{\phi}}_1 \\ 
		-\hat{\bm{\phi}}_1^* & -\hat{\bm{\phi}}_2^*
	\end{pmatrix} 
  \nonumber\\
=&  \pm
    \begin{pmatrix} 
	- \hat{\bm{\phi}}_1^* \hat{\bm{\phi}}_1 + \hat{\bm{\phi}}_2^* \hat{\bm{\phi}}_2  &	- 2\hat{\bm{\phi}}_1 \hat{\bm{\phi}}_2^* \\ 
		-2\hat{\bm{\phi}}_1^* \hat{\bm{\phi}}_2 & \hat{\bm{\phi}}_1^* \hat{\bm{\phi}}_1 - \hat{\bm{\phi}}_2^* \hat{\bm{\phi}}_2
	\end{pmatrix}
 .
\end{align}


The color direction field is indeed normalized:
\begin{align}
  n_A(x) n_A(x) = 1
  .
\end{align}
Using the matrix form, this is shown as
\begin{align}
 n_A n_A =& \frac12 {\rm tr}[n_A \sigma_A n_B \sigma_B ] = \frac12 {\rm tr}[\hat\Theta \sigma_3 \hat\Theta^{\dagger} \hat\Theta \sigma_3 \hat\Theta^{\dagger}] 
\nonumber\\
=& \frac12 {\rm tr}[ \sigma_3 \sigma_3 ]  = \frac12{\rm tr}[ \bm{1} ] =  1 
 ,
\end{align}
where we have used ${\rm tr}[\sigma_A  \sigma_B]=2\delta_{AB}$.


Under the gauge transformation (\ref{gauge-transf}), the color field  $\bm{n}(x) = n_A(x) \sigma_A$ defined in this way transforms according to the adjoint representation:
\begin{align}
 \bm{n}(x) \to \bm{n}^\prime(x) = U(x) \bm{n}(x) U^\dagger(x)    
  ,
  \label{W-4a}
\end{align}
provided that the gauge transformation of the scalar field obeys 
\begin{align}
& \Phi(x) \to \Phi^\prime(x) = U(x) \Phi(x) 
\nonumber\\
  \Longrightarrow & \Theta(x) \to \Theta^\prime(x) = U(x) \Theta(x)  .
  \label{W-4b}
\end{align}
Notice that the color field could be alternatively defined by
$\bm{n}(x)= \Theta^{\dagger}(x)  \sigma_3 \Theta(x)$. But the gauge transformation property (\ref{W-4a}) is lost by this choice of the color field.

\subsection{The residual field in terms of the color field}

We show that the color direction field $\bm{n}$ is used to represent the residual field $\mathscr{V}$. 
Notice that 
\begin{align}
& {D}_{\mu}[\mathscr{V}] \hat{\Theta}(x)  = 0 \Longrightarrow \mathscr{D}_{\mu}[\mathscr{V}] \bm{n}(x) = 0
 ,
 \label{def-V-step2}
\end{align}
which follows easily from the Leibniz rule for the covariant derivative as 
\begin{align}
  \mathscr{D}_{\mu}[\mathscr{V}] \bm{n}
=& \mathscr{D}_{\mu}[\mathscr{V}] (\hat{\Theta} \sigma_3 \hat{\Theta} ^\dagger)
\nonumber\\
=& ( {D}_{\mu}[\mathscr{V}] \hat{\Theta} ) \sigma_3 \hat{\Theta} ^\dagger + \hat{\Theta} \sigma_3 ({D}_{\mu}[\mathscr{V}] \hat{\Theta} )^\dagger
 .
\end{align}
The converse is not necessarily true, since the condition ${D}_{\mu}[\mathscr{V}] \hat{\Theta}  = 0$ is stronger than $\mathscr{D}_{\mu}[\mathscr{V}] \bm{n} = 0$.

Thus we find the relationship from (\ref{def-V-step1}) and (\ref{def-V-step2}):
\begin{align}
& {D}_{\mu}[\mathscr{V}] \hat{\Phi}(x)  = 0 \Longrightarrow  {D}_{\mu}[\mathscr{V}] \hat{\Theta}(x) = 0
 \Longrightarrow \mathscr{D}_{\mu}[\mathscr{V}] \bm{n}(x) = 0
 .
\end{align}

Therefore, $\mathscr{V}$ is also expressed in terms of $\bm{n}$ by solving 
\begin{align}
 \mathscr{D}_{\mu}[\mathscr{V}] \bm{n}(x) = 0 
 \Longleftrightarrow 
 \partial_\mu \bm{n}(x) -ig [\mathscr{V}_\mu(x) , \bm{n}(x)] = 0 
 ,
\end{align}
which is also written in the vector form:
\begin{align}
 \mathscr{D}_{\mu}[\mathbf{V}] \mathbf{n}(x) = 0 
 \Longleftrightarrow 
 \partial_\mu \mathbf{n}(x) + g \mathbf{V}_\mu(x) \times \mathbf{n}(x) = 0 
 .
\end{align}
By taking the commutator with $\bm{n}$, we have
\begin{align}
 ig^{-1}[ \bm{n}(x) ,  \partial_\mu \bm{n}(x)] =&  [ \bm{n}(x) , [\bm{n}(x), \mathscr{V}_\mu(x) ]]   
 .
\end{align}
By using the formula: for any $su(2)$ valued function $\mathscr{F}$, 
\begin{align}
\mathscr{F}=\bm{n} (\bm{n}  \cdot \mathscr{F} )+[ \bm{n} , [\bm{n} , \mathscr{F}   ]]  , 
\end{align}
with the definition,
\begin{align}
(\bm{n}  \cdot \mathscr{F} )  
:= {\rm tr}[\bm{n} \mathscr{F} ] 
=  {n}^A \mathscr{F}^A 
 ,
\end{align}
see e.g., \cite{KKSS15} for the proof, 
$\mathscr{V}$ is also expressed in terms of $\bm{n}$:
\begin{align}
\mathscr{V}_\mu(x) =& \bm{n}(x)(\bm{n}(x) \cdot \mathscr{V}_\mu(x)) + ig^{-1}[ \bm{n}(x) ,  \partial_\mu \bm{n}(x)] 
 .
\end{align}

By introducing $v_\mu$ by
\begin{align}
v_\mu(x) :=  (\bm{n}(x) \cdot \mathscr{V}_\mu(x)) 
= {\rm tr}[\bm{n}(x) \mathscr{V}_\mu(x)] 
=  {n}^A(x) \mathscr{V}_\mu^A(x)
 ,
\end{align}
we have
\begin{align}
\mathscr{V}_\mu(x) =& v_\mu(x) \bm{n}(x) + ig^{-1}[ \bm{n}(x) ,  \partial_\mu \bm{n}(x)] 
 ,
 \label{V-h-n}
\end{align}
which is written in the vector notation as 
\begin{align}
    \mathbf{V}_{\mu}(x)   
=  v_\mu(x) \mathbf{n}(x)   - g^{-1}  \mathbf{n}(x) \times \partial_\mu \mathbf{n}(x) 
 ,
 \label{V-h-n2}
\end{align}
with the components, 
\begin{align}
    \mathscr{V}_{\mu}^A(x)  
=  v_\mu(x) n_A(x)  - g^{-1} \epsilon^{ABC} n_B(x)  \partial_\mu n_C(x)
  .
 \label{V-h-n3}
\end{align}
We find that $v_\mu$ is expressed in terms of the scalar field $\Phi$ as 
\begin{align}
    v_\mu(x)  =& \mp  g^{-1} i[ \partial_\mu \hat{\Phi}(x)^\dagger   \hat{\Phi}(x)   - \hat{\Phi}(x)^\dagger  \partial_\mu \hat{\Phi}(x)   ]
\nonumber\\
=& \pm 2 g^{-1}  i\hat{\Phi}(x)^\dagger \partial_\mu \hat{\Phi}(x)  
=  \mp  2 g^{-1} i\partial_\mu \hat{\Phi}(x)^\dagger   \hat{\Phi}(x)
  .
 \label{h-def}
\end{align}
In fact, the parallel component of $\mathscr{V}_\mu$ (\ref{V-h-n}) is extracted by using the representation (\ref{def-color-f2}) of $\bm{n}$  and the formula (\ref{V-SU2-1}) for $\mathscr{V}_\mu$ as 
\begin{align}
  v_{\mu}  
 =&  {\rm tr}[\bm{n} \mathscr{V}_\mu ] 
\nonumber\\
 =& g^{-1} {\rm tr}[\pm \hat\Theta \sigma_3 \hat\Theta^{\dagger}  (-i)  \partial_{\mu}\hat{\Theta}    \hat{\Theta} ^\dagger] 
\nonumber\\
=&  \mp   g^{-1} i {\rm tr} [\sigma_3 \hat{\Theta} ^\dagger \partial_{\mu} \hat{\Theta} ]
\nonumber\\
=&  \mp   g^{-1} i (\hat{\tilde\Phi}^\dagger \partial_{\mu} \hat{\tilde\Phi} - \hat{\Phi}^\dagger \partial_\mu \hat{\Phi})  
\nonumber\\
=& \mp   g^{-1} i ( \hat{\Phi}^t   \partial_{\mu}  \hat{\Phi}^* - \hat{\Phi}^\dagger \partial_\mu \hat{\Phi})  
\nonumber\\
=& \mp  2  g^{-1}  i  \hat{\Phi}^t   \partial_{\mu}  \hat{\Phi}^*   
= \pm  2  g^{-1}  i   \hat{\Phi}^\dagger \partial_\mu \hat{\Phi}  
  ,
\end{align}
where we have used
$
\hat{\tilde\Phi}^\dagger \partial_{\mu} \hat{\tilde\Phi}
= \hat{\Phi}^t \epsilon^t \epsilon \partial_{\mu}  \hat{\Phi}^*
= \hat{\Phi}^t   \partial_{\mu}  \hat{\Phi}^*
$,
$
\hat{\Phi}^t   \partial_{\mu}  \hat{\Phi}^*
=(\partial_{\mu}  \hat{\Phi}^\dagger \hat{\Phi}    )^t
= \partial_{\mu}  \hat{\Phi}^\dagger \hat{\Phi}  
$
and
$\hat{\Phi}^\dagger\hat{\Phi}=1$.

The residual field $\mathscr{V}$ is decomposed into the two parts which are parallel $\mathscr{V}^\parallel$ and perpendicular $\mathscr{V}^\perp$ to the color direction field 
\begin{align}
\mathscr{V}_{\mu}(x)=\mathscr{V}^\parallel_{\mu} (x) + \mathscr{V}^\perp_{\mu} (x) 
 .
\end{align}
The parallel part is obtained as 
\begin{align}
   g \mathscr{V}^\parallel_{\mu}(x) 
= g \mathscr{V}_{\mu}^A(x) n_A(x) \bm{n}(x)
=  v_\mu(x) \bm{n}(x)
  ,
  \label{Vpa-a1}
\end{align}
which has the vector notation,
\begin{align}
 g\mathbf{V}_{\mu}^\parallel(x) = (g \mathbf{V}_{\mu}(x) \cdot \mathbf{n}(x))\mathbf{n}(x)  =
  v_\mu(x) \mathbf{n}(x)    
   .
  \label{Vpa-v2}
\end{align}
The perpendicular part is obtained 
\begin{align}
   g \mathscr{V}_{\mu}^\perp(x)   
 =      i  [ \bm{n}(x),   \partial_\mu \bm{n}(x)]      
  ,
\end{align}
which has the vector notation,
\begin{align}
 g\mathbf{V}_{\mu}^\perp(x)
  =  -  \mathbf{n}(x) \times \partial_\mu \mathbf{n}(x) 
  =   \partial_\mu \mathbf{n}(x) \times \mathbf{n}(x) 
  .
  \label{V-3b2}
\end{align}
Notice that
$
 g\mathbf{V}_{\mu}^\perp
 = g\mathbb{V}_{\mu} \times \mathbf{n}
 =  \mathbf{n} \times (\mathbf{n} \times \partial_\mu \mathbf{n} )
  .
$

It should be remarked that the massive field $\mathscr{W}_{\mu}$ for the fundamental scalar case is not perpendicular to the color direction field $\bm{n}$,
\begin{align}
 \mathscr{W}_{\mu}(x)  \cdot \bm{n}(x)  \not= 0 ,
\end{align}
which is sharp contrast to the adjoint scalar case 
$\mathscr{W}_{\mu}(x)  \cdot \bm{n}(x) = 0$.
Under the gauge transformation, $v_\mu$ defined by (\ref{h-def}) transforms 
\begin{align}
 v_\mu(x) \to  v_\mu^\prime(x) 
=& v_\mu(x)   \mp i2 g^{-1}  \hat{\Phi}(x)^\dagger U(x)^\dagger \partial_\mu U(x) \hat{\Phi}(x)  
\nonumber\\
=& v_\mu(x)   \pm i2 g^{-1}  \hat{\Phi}(x)^\dagger \partial_\mu U(x)^\dagger  U(x) \hat{\Phi}(x)  
  .
\end{align}
This result is also obtained from (\ref{V-h-n}).

\subsection{Field strength in terms of the color field}

According to the decomposition of the gauge field $\mathscr{A}_\mu(x)=\mathscr{V}_\mu(x)+\mathscr{W}_\mu(x)$, the field strength $\mathscr{F}_{\mu\nu}(x)$ of the original gauge field $\mathscr{A}_\mu(x)$ is decomposed as  
\begin{align}
  \mathscr{F}_{\mu\nu}[\mathscr{A}] :=& \partial_\mu \mathscr{A}_\nu - \partial_\nu \mathscr{A}_\mu -i g [ \mathscr{A}_\mu , \mathscr{A}_\nu ]
\nonumber\\
=& \mathscr{F}_{\mu\nu}[\mathscr{V}+\mathscr{W}]
\nonumber\\
=& \mathscr{F}_{\mu\nu}[\mathscr{V}]  + \mathscr{D}_\mu[\mathscr{V}] \mathscr{W}_\nu - \mathscr{D}_\nu[\mathscr{V}] \mathscr{W}_\mu 
-i g [ \mathscr{W}_\mu , \mathscr{W}_\nu ] .
\label{F-decomposition1}
\end{align}
where $\mathscr{F}_{\mu\nu}[\mathscr{V}]$ is the field strength of the residual gauge field $\mathscr{V}$ defined by 
\begin{align}
   \mathscr{F}_{\mu\nu}[\mathscr{V}]
 :=& \partial_\mu \mathscr{V}_\nu - \partial_\nu \mathscr{V}_\mu - ig [\mathscr{V}_\mu , \mathscr{V}_\nu ]
 ,
\end{align}
and  $\mathscr{D}_\mu[\mathscr{V}]$ is the covariant derivative in the background gauge field $\mathscr{V}_\mu$.

By substituting the decomposition (\ref{F-decomposition1}) of the field strength into the $SU(2)$ gauge-scalar Lagrangian with a radially fixed fundamental scalar field, we obtain the decomposition of the Yang-Mills Lagrangian density with the mass term $\frac{1}{2 } M_{W}^2 \mathscr{W}^\mu  \cdot \mathscr{W}_\mu$ generated by the gauge-invariant BEH mechanism,
\begin{align}
   \mathscr{L}_{\rm RF} 
=& -\frac{1}{4} \mathscr{F}_{\mu\nu}[\mathscr{V}+\mathscr{W}] \cdot \mathscr{F}^{\mu\nu}[\mathscr{V}+\mathscr{W}]
+ \frac{1}{2 } M_{W}^2 \mathscr{W}^\mu  \cdot \mathscr{W}_\mu 
\nonumber\\ 
=& -\frac{1}{4} \mathscr{F}_{\mu\nu}[\mathscr{V}] \cdot \mathscr{F}^{\mu\nu}[\mathscr{V}]
\nonumber\\&
- \frac{1}{4} (\mathscr{D}_\mu[\mathscr{V}] \mathscr{W}_\nu - \mathscr{D}_\nu[\mathscr{V}] \mathscr{W}_\mu)^2
+ \frac{1}{2 } M_{W}^2 \mathscr{W}^\mu  \cdot \mathscr{W}_\mu 
\nonumber\\&
+ \frac{1}{2}  \mathscr{F}_{\mu\nu}[\mathscr{V}] \cdot ig[ \mathscr{W}^\mu , \mathscr{W}^\nu ]
\nonumber\\&
+ \frac{1}{2}  (\mathscr{D}_\mu[\mathscr{V}] \mathscr{W}_\nu - \mathscr{D}_\nu[\mathscr{V}] \mathscr{W}_\mu) \cdot ig[ \mathscr{W}^\mu , \mathscr{W}^\nu ]
\nonumber\\&
- \frac14 (i g [ \mathscr{W}_\mu , \mathscr{W}_\nu ])^2
 ,
\label{L-VW1}
\end{align}
where each term is $SU(2)$ invariant and the term $- \frac{1}{2} \mathscr{F}^{\mu\nu}[\mathscr{V}] \cdot (\mathscr{D}_\mu[\mathscr{V}] \mathscr{W}_\nu - \mathscr{D}_\nu[\mathscr{V}] \mathscr{W}_\mu)
$ linear in $\mathscr{W}^\mu$ is eliminated. 
\footnote{
In the case of the adjoint scalar field, it is shown that 
$\mathscr{F}_{\mu\nu}[\mathscr{V}](x)$ and $-i g [ \mathscr{W}_\mu(x) , \mathscr{W}_\nu(x) ] $ are parallel to the color direction field $\bm{n}(x)=\hat{\bm{\phi}}(x)$, 
while $\mathscr{D}_\mu[\mathscr{V}] \mathscr{W}_\nu(x) - \mathscr{D}_\nu[\mathscr{V}] \mathscr{W}_\mu(x)$
is perpendicular to  $\bm{n}(x)=\hat{\bm{\phi}}(x)$, which follows from  the defining equations: 
 $\bm{n}(x) \cdot \mathscr{W}_\mu(x)=0$. 
 Therefore, the terms $- \frac{1}{2} \mathscr{F}^{\mu\nu}[\mathscr{V}] \cdot (\mathscr{D}_\mu[\mathscr{V}] \mathscr{W}_\nu - \mathscr{D}_\nu[\mathscr{V}] \mathscr{W}_\mu)
$ linear in $\mathscr{W}^\mu$ and $\frac{1}{2}  (\mathscr{D}_\mu[\mathscr{V}] \mathscr{W}_\nu - \mathscr{D}_\nu[\mathscr{V}] \mathscr{W}_\mu) \cdot ig[ \mathscr{W}^\mu , \mathscr{W}^\nu ]$ vanish and do not appear in the Lagrangian. 
}
Here, $\mathscr{V}_\mu$ is supposed to be given by the pure gauge form (\ref{V-SU2-1}) written in terms of the normalized scalar field, i.e., gauge group element $\hat\Theta \in SU(2)$ and $\mathscr{W}_\mu$ is regarded as the independent fundamental field variables representing massive vector boson fields. 
We observe that the vector field $\mathscr{W}_\mu$ has the ordinary kinetic term and the mass term (in the tree level).  
Therefore, there is a massive vector pole in the propagator of $\mathscr{W}_\mu$ (after a certain gauge fixing). Thus, $\mathscr{W}_\mu$ is not an auxiliary field, but is a propagating field with the mass $M_{W}$  (up to possible quantum corrections).

In the case of the fundamental scalar field, the residual gauge mode $\mathscr{V}_\mu$ has the pure gauge form (\ref{V-SU2-1}) $\mathscr{V}_\mu=ig^{-1} \hat{\Theta}  \partial_{\mu}  \hat{\Theta} ^\dagger$. Therefore, the field strength $\mathscr{F}_{\mu\nu}[\mathscr{V}]$ of $\mathscr{V}_\mu$ vanishes except for the singular points at which the commutator of the partial derivatives does not commute:  
\begin{align}
   \mathscr{F}_{\mu\nu}[\mathscr{V}](x)
=&   -i g^{-1} [\partial_\mu, \partial_\nu] \hat\Theta(x)  \hat\Theta(x)^\dagger
\nonumber\\
=&   i g^{-1} \hat\Theta(x) [\partial_\mu, \partial_\nu] \hat\Theta(x)^\dagger
 .
 \label{F-V-adj}
\end{align}

The result (\ref{F-V-adj}) should be obtained by using the residual gauge mode (\ref{V-h-n}) written in terms of the color direction field constructed from the normalized scalar field, 
\begin{align}
\mathscr{V}_\mu(x) =& v_\mu(x) \bm{n}(x) + ig^{-1}[ \bm{n}(x) ,  \partial_\mu \bm{n}(x)] 
 ,
\end{align}
By using the vector form (\ref{V-h-n2}) or the component form (\ref{V-h-n3}), the direct calculations lead to  
\begin{align}
   \mathbf{F}_{\mu\nu}[\mathbf{V}](x)
 =& \mathbf{n}(x)[\partial_\mu v_\nu(x) - \partial_\nu v_\mu(x) 
\nonumber\\ &
- g^{-1} \mathbf{n}(x) \cdot (\partial_\mu \mathbf{n}(x) \times \partial_\nu \mathbf{n}(x))]  
\nonumber\\ &
 + g^{-1} [\partial_\mu ,  \partial_\nu ] \mathbf{n}(x) \times \mathbf{n}(x) 
 ,
\end{align}
which has the Lie-algebra valued form,
\begin{align}
 \mathscr{F}_{\mu\nu}[\mathscr{V}](x)
 =& \bm{n}(x) \{ \partial_\mu v_\nu(x) - \partial_\nu v_\mu(x) 
\nonumber\\ &
+ ig^{-1} \bm{n}(x) \cdot [\partial_\mu \bm{n}(x) ,  \partial_\nu \bm{n}(x)] \}  
\nonumber\\ &
 - ig^{-1} [ [\partial_\mu ,  \partial_\nu ] \bm{n}(x) , \bm{n}(x) ]
 .
\label{F-V-adj-n}
\end{align}
In these calculations we have only used a fact that the color direction field has the unit length $\mathbf{n}(x) \cdot \mathbf{n}(x)=1$. 
From
\begin{align}
  [\partial_\mu ,  \partial_\nu ] \bm{n}(x) 
 =& [\partial_\mu ,  \partial_\nu ] (\hat\Theta(x) \sigma_3 \hat\Theta(x)^{\dagger})
\nonumber\\ 
 =& [\partial_\mu ,  \partial_\nu ] \hat\Theta(x) \sigma_3 \hat\Theta(x)^{\dagger} + \hat\Theta(x) \sigma_3 [\partial_\mu ,  \partial_\nu ]  \hat\Theta(x)^{\dagger}
 ,
\end{align}
 we find that the last term of (\ref{F-V-adj-n}) corresponds to (\ref{F-V-adj}):
\begin{align}
 [\partial_\mu ,  \partial_\nu ] \hat\Theta(x) = 0 
 \Longrightarrow 
 [\partial_\mu ,  \partial_\nu ] \bm{n}(x)  = 0 
 .
\end{align}
This suggest that the Abelian-like field strength identically vanishes in the fundamental scalar case:
\begin{align}
    F_{\mu\nu} (x)
 :=&   \partial_\mu v_\nu(x) - \partial_\nu v_\mu(x) 
\nonumber\\ 
  &+ ig^{-1} \bm{n}(x) \cdot [\partial_\mu \bm{n}(x) ,  \partial_\nu \bm{n}(x)]  \equiv 0 
 .
\label{f-V-adj-n}
\end{align}
where $v_\mu$ is given by (\ref{h-def}) and $\bm{n}$ is given by (\ref{def-color-f2}). 
Indeed, this identity is derived by representing the normalized scalar field $\hat\Phi$ in terms of the three angles for the three sphere $S^3$. 
Therefore, we obtain the formula for the field strength of the residual model written in terms of the color direction field:
\begin{align}
  \mathscr{F}_{\mu\nu}[\mathscr{V}](x)
 = - ig^{-1} [ [\partial_\mu ,  \partial_\nu ] \bm{n}(x) , \bm{n}(x) ]
 .
\label{F-V-adj-nf}
\end{align}
This result is consistent with the fact that there are no residual massless gauge fields in the fundamental scalar case after the BEH mechanism takes place.


\section{Conclusion and discussion}

We have extended a gauge-independent description \cite{Kondo16} of the BEH or Higgs mechanism \cite{Higgs1,Higgs2,Higgs3} by which massless gauge bosons acquire their mass to include a fundamental scalar field.
Consequently, we can introduce a gauge-invariant mass term in the Yang-Mills theory.

The conventional description of the BEH mechanism  requires a non-vanishing vacuum expectation value of the scalar field $\langle 0| \phi(x) |0 \rangle=v$,  which is clearly gauge dependent and impossible to be realized without fixing the gauge.  
In the new description \cite{Kondo16}, instead, the scalar field is supposed to obey a  gauge-invariant condition which forces the radial length of the scalar field to have a certain fixed value $||\phi(x)||=v$ without breaking the gauge symmetry. 
Therefore, this extension enables one to study quark confinement and mass gap in the pure Yang-Mills theory as the implications of the BEH mechanism in the ``complementary'' gauge-scalar model, as suggested from the Fradkin--Shenker continuity \cite{FS79} and the Osterwalder--Seiler theorem \cite{OS78} on the lattice.

The new description allows one to decompose the original gauge field $\mathscr{A}$ into the massive vector mode $\mathscr{W}$ and the residual gauge mode $\mathscr{V}$, $\mathscr{A}=\mathscr{W}+\mathscr{V}$ in the gauge-independent way. 
The massive  vector mode $\mathscr{W}_\mu$ will rapidly fall off in the distance and hence it is identified with the short-distance (or high-energy) mode. 
Therefore,  massive vector modes $\mathscr{W}$ mediate only the short-range force between quark sources.  
Consequently, the long-range force giving a linear piece of the static quark potential responsible for quark confinement must be mediated by the residual gauge mode $\mathscr{V}$.
In the case of the adjoint scalar field,  the residual gauge mode include massless gauge boson which is able to mediate the long-range force. 
In the case of the fundamental scalar field, there are no massless gauge bosons in the residual mode $\mathscr{V}$ once the BEH mechanism occurs.  In fact,  the residual gauge mode $\mathscr{V}$ has exactly the same form as the pure gauge $\mathscr{V}=ig^{-1}UdU^{-1}$ with the group element $U$ which is written in terms of the scalar field $\Phi$ alone. Therefore, solitons and defects converging to the pure gauge in the long distance could be dominant field configurations responsible  for quark confinement.

For quark confinement, the two cases give the different perspective.  
In the adjoint scalar field case ($SU(2) \to U(1)$),
the external quark source in the fundamental representation  cannot be screened by the adjoint scalar field and the chromoelectric flux connecting a pair of quark and antiquark is formed for any distance $r$ larger than a certain distance $r_0$, $r > r_0$, while for $r< r_0$ the Coulomb-like perturbative part becomes dominant. 
In the fundamental scalar field case  ($SU(2) \to \{ 0 \}$),
the external quark source in the fundamental representation  can be screened by the fundamental scalar field and the chromoelectric flux connecting a pair of quark and antiquark will break at certain distance $r =r_c \simeq 2m/\sigma$ with the mass $m$ of the scalar particle and the string tension $\sigma$. 
The static quark potential exhibits the linear potential in the intermediate region $r_0 < r < r_c$, and flattens in the long-distance region $r > r_c$. 
This situation is similar to the realistic QCD in which light dynamical quarks are included into the theory. 
For gluon confinement, we can calculate gluon propagators leading to positivity violation, which is consistent with gluon confinement as shown in a subsequent paper \cite{Kondo-etal18}.


Based on the general framework given in this paper, we will demonstrate its validity in understanding confinement for various choice of the spacetime dimension $D=2,3,4$ and the gauge group $G=U(1)$, $SU(2)$, $SU(2) \times U(1)$, $SU(3)$  in subsequent papers where  
the detailed analyses on the solution of the field equations of the complementary gauge-scalar model will be given.

{\it Acknowledgements}\ ---
This work is  supported by Grant-in-Aid for Scientific Research, JSPS KAKENHI Grant Number (C) No.15K05042.

\appendix
\section{$U(1)$ gauge-scalar model}

The instanton configuration in $D=2$ dimensional Euclidean space-time can be identified with the static vortex (line-like defect) in $D=2+1$ dimensional Minkowski space-time. 
The finite-action configurations for $D=2$ spacetime are characterized by an integer $N$, the \textbf{winding number}, just as they are for four-dimensional gauge field theories. 
Usually, the winding number is obtained as the integral over the circle $C_{\infty}$ located at infinity:  
\begin{align}
 N := \frac{1}{2\pi} \oint_{C_{\infty}} dx^{a} A_{a}  \quad (a=1,2)
,
\end{align}
and is equivalently written using the Stokes theorem as the area integral over the  surface $S_{\infty}$ enclosed by the loop $C_{\infty}$:
\begin{align}
 N := \frac{1}{4\pi} \int_{S_{\infty}: \partial S_{\infty}=C_{\infty}} d^2x \epsilon^{ab} F_{ab}[A]  \quad (a,b=1,2)
 ,
\end{align}
where $F_{ab}[A]:=\partial_a A_b - \partial_b A_a$ is the field strength of the field $A_a$. 
This guarantees that the winding number is gauge independent, namely, does not depend on the gauge choice. 
The winding number is equal to the first Chern number of the magnetic field.

According to the procedure presented in this paper, however, we can explicitly separate the original gauge field $A_\mu$ into the gauge-invariant massive modes $W_\mu$ and the residual field $V_\mu$.
Remarkably, an arbitrary circle $C$ can be used to give the winding number by choosing $V_\mu$ as far as it encloses the center of the defect: 
\begin{align}
 N = \frac{1}{2\pi} \oint_{C} dx^{a} V_{a}  
 = \frac{1}{4\pi} \int_{S: \partial S=C} d^2x \epsilon^{ab} F_{ab}[V]  
 ,
\end{align}
where $F_{ab}[V]:=\partial_a V_b - \partial_b V_a$ is the field strength of the residual field $V_a$ and $S$ is an arbitrary surface whose boundary is equal to the loop $C$. 
This follows from the observation that in the long-distance $\rho \to \infty$, the massive mode $W_\mu$ does not contribute to the integral and the residual mode $V_\mu$ alone survives the limit $\rho \to \infty$. 

We find that $N$ takes an integral value. For $\hat\phi(x)=e^{ig\varphi(x)} \in U(1)$, we have  
$
 V_{a}(x)  =  \partial_a \varphi(x) 
$ and 
\begin{align}
 N =& \frac{1}{2\pi} \oint_{C} dx^{a} \partial_a \varphi(x) 
= \frac{1}{2\pi} \oint_{C} d \varphi(x) 
\nonumber\\  
=& \frac{1}{2\pi}(\varphi(2\pi) - \varphi(0))
 \in \mathbb{Z} ,
\end{align}
since $\varphi(x)$ must have the property $\varphi(2\pi) - \varphi(0)=2\pi n$ due to single-valuedness of $\hat\phi$. 


The vector field $\bm{V}=(V_1,V_2)$ has vanishing divergence and rotation except on the singular points:
$
 \nabla \cdot \bm{V} = 0
$, 
$
 \nabla \times \bm{V} = \bm{0}
$.
Therefore, we have
\begin{align}
 N =& \frac{1}{2\pi} \oint_{C} d\bm{r} \cdot \bm{V} 
=  \frac{1}{2\pi} \oint_{C} (dx^{1} V_{1} + dx^{2} V_{2} )
\nonumber\\  
=&  \frac{n}{2\pi}  \oint_{C} \left( dx^{1} \frac{-x_2}{x_1^2+x_2^2} + dx^{2} \frac{ x_1}{x_1^2+x_2^2} \right)  
\nonumber\\  
 =&  \frac{n}{2\pi} \oint_{C} d \arctan \frac{x_2}{x_1}   
\nonumber\\  
 =&  \frac{n}{2\pi} \oint_{C} d \varphi =  n
 .
\end{align}
This means that the vector field $\bm{V}$ is equal to the gradient of the angle $\varphi$:
\begin{align}
 \bm{V}(x_1,x_2) = n \nabla \varphi(x_1,x_2), \ \varphi(x_1,x_2) =  \arctan \frac{x_2}{x_1} .
\end{align}
The rotation of the vector field $V$ in the whole space is given by
\begin{align}
 \nabla \times \bm{V}(x_1,x_2) = 2\pi n \delta(x_1)\delta(x_2) \bm{e}_3 , 
\end{align}
which follows from
\begin{align}
 N =&  \frac{1}{2\pi} \int_{S: \partial S=C} d \bm{S}  \cdot  (\nabla \times \bm{V})
\nonumber\\ 
=&  \frac{1}{2\pi} \int_{S: \partial S=C} dx_1 dx_2 \bm{e}_3 \cdot  2\pi n \delta^2(x_1,x_2) \bm{e}_3 = n 
 .
\end{align}


We discuss explicitly the $D=3$ case. 
For $D=2+1$, the vortex is the relevant configuration. 
In what follows, especially, we pay attention to the difference between the usual Nielsen-Olesen vortex in the $U(1)$ gauge-Higgs model and the new vortex in the present $U(1)$ gauge-scalar model with a radially fixed scalar field. 

\par
We adopt the static (i.e., time-independent) and axially symmetric Ansatz:
\begin{align}
 \phi (x) =& F(\rho) e^{in \varphi} , \ n \in  \mathbb{Z} .
\nonumber\\ 
 A_{0}(x) =& 0 , 
\nonumber\\  
 A_a(x) =& - \epsilon_{ab} \frac{x_b}{\rho} A_\varphi(\rho)  = - \epsilon_{ab} \frac{x_b}{\rho} \frac{n  }{q } \frac{ a(\rho)}{ \rho} \ (a,b=1,2) , 
\label{U1-Ansatz1}
\end{align}
where 
\begin{align}
 \varphi = \arctan \frac{x_2}{x_1} , \ \rho := \sqrt{x_1^2+x_2^2} .
\end{align}
Under this Ansatz,  the field equations (\ref{CP-AH-fe})  are reduced to coupled nonlinear ordinary differential equations  for $A$ and $F$ as functions of $\rho$:%
\begin{align}
0 =&- \frac{d}{d\rho} \left\{ \frac{1}{\rho} \frac{d}{d\rho} [ \rho A(\rho)] \right\}
+ 2q^2F^{2}(\rho) \biggl[  A(\rho) - \frac{n }{q\rho} \biggr]    ,
\label{CP-AH-fe2b}
\\
0 =&- \frac{1}{\rho} \frac{d}{d\rho} \biggl[ \rho \frac{d}{d\rho} F(\rho) \biggr]
+  q^2 \biggl[ \frac{n}{q\rho} -  A(\rho) \biggr]^{2} F(\rho)
\nonumber\\ &
 + \lambda \biggl[ F^{2}(\rho) - \frac{\mu^{2}}{\lambda} \biggr] F(\rho)   . 
\label{CP-AH-fe2a}
\end{align}
For the radially fixed case, the field equations are given by
\begin{align}
0=&  F(\rho)^2 - \frac{1}{2} v^2   . 
\label{fixed-const}
\\
0=& - \frac{1}{\rho} \frac{d}{d\rho} \biggl[ \rho \frac{d}{d\rho} F(\rho) \biggr]
+  q^2 \biggl[ \frac{n}{q\rho} -  A(\rho) \biggr]^{2} F(\rho) - u(\rho)  F(\rho) , 
\label{CP-AH2-fe2a}
\\
0=& - \frac{d}{d\rho} \left\{ \frac{1}{\rho} \frac{d}{d\rho} [ \rho A(\rho)] \right\}
+ 2q^2F^{2}(\rho) \biggl[  A(\rho) - \frac{n }{q\rho} \biggr] .
\label{CP-AH2-fe2b}
\end{align}
For arbitrary values of $\lambda$ and $q$, the explicit analytical solutions for (\ref{CP-AH-fe2b}) and (\ref{CP-AH-fe2a}) are not known. 
In the limit $\lambda=\infty$, especially, the magnitude of the scalar field is fixed $F(\rho)=\frac{v}{\sqrt{2}}$, and therefore $A$ is solved:
\begin{align}
 F(\rho) = \frac{v}{\sqrt{2}} , \quad A(\rho) = \frac{n }{q\rho} ,
\end{align}
which gives the vacuum solution with the lowest energy $E=0$. 
The soliton solution with a finite energy approaches the vacuum solution in the  large $\rho$ asymptotic region.

In the radially fixed scalar case, $F(\rho)=\frac{v}{\sqrt{2}}$ for any $\rho$, but the gauge field is not restricted to the vacuum solution and can deviate from it. By solving (\ref{fixed-const}), the equations (\ref{CP-AH2-fe2a}) and (\ref{CP-AH2-fe2a}) reduce to 
\begin{align}
 & u(\rho) = q^2 \biggl[ \frac{n}{q\rho} -  A(\rho) \biggr]^{2} , 
\label{Lag-multi}
\\
0=&- \frac{d}{d\rho} \left\{ \frac{1}{\rho} \frac{d}{d\rho} [ \rho A(\rho)] \right\}
+  q^2v^2 \biggl[  A(\rho) - \frac{n }{q\rho} \biggr] ,
\label{eq-A}
\end{align}
the first equation (\ref{Lag-multi}) determines the Lagrange multiplier field $u$ once the solution of the second equation (\ref{eq-A}) for the gauge field $A_\mu$ is obtained. The second equation (\ref{eq-A}) can be solved by using the ansatz:
\begin{align}
 A_a(x) 
=& - \epsilon_{ab} \frac{x_b}{\rho} \frac{n}{q} \frac{ 1-\rho w(\rho)}{ \rho} \ (a,b=1,2)  , 
\label{CP-Ansatz1}
\end{align}
where $w$ must satisfy 
\begin{align}
& \frac{d^2 w(\rho)}{d\rho^2} + \frac{1}{\rho} \frac{d w(\rho)}{d\rho} 
- \left( M_{W}^2 + \frac{1}{\rho^2} \right) w (\rho)  =  0 .
\end{align}
The equation for $w$ is the modified Bessel differential equation and the solution is given by the modified Bessel functions. The two linearly independent solutions are denoted by  $I_1(M_{W}\rho)$ and $K_1(M_{W}\rho)$. 
The solution must be determined so as to satisfy the boundary conditions.

The energy of the field configuration satisfying the field equation is obtained by the variation of the Hamiltonian:
\begin{equation}
H = \int d^{3} x \left\{ \frac{1}{2}   \bm{B}^{2} + \frac{1}{2}  \bm{E}^{2}  
+   | D_{k} \phi |^{2} +   | D_{0} \phi |^{2} + V (\phi) \right\} .
\label{AH-E}
\end{equation}
By substituting the Ansatz (\ref{U1-Ansatz1}) into (\ref{AH-E}), we have
\begin{align}
E =& \int^{+\infty}_{-\infty} dz \int^{+\infty}_{-\infty} 2\pi \rho d \rho
\left\{ \frac{1}{2\rho^{2}} \biggl[ \frac{d}{d\rho} ( \rho A(\rho) ) \biggr]^{2}
+ \biggl[ \frac{d}{d\rho} F(\rho) \biggr]^{2} \right. \nonumber\\
& \left. + q^2F^{2}(\rho) \biggl[   A(\rho) - \frac{n}{q\rho} \biggr]^{2} 
+ \frac{\lambda}{2} \left( F^{2}(\rho) - \frac{\mu^{2}}{\lambda} \right)^{2} \right\} \geq 0 .
\label{C23-AH-energy}
\end{align}
In fact, the Euler-Lagrange equation for $A$ and $F$ agree with (\ref{CP-AH-fe2b}) and (\ref{CP-AH-fe2a}).

For the energy of a vortex per unit length to be finite, the profile functions must satisfy the boundary condition at $\rho=\infty$:
\begin{align}
&  F(\rho)  \simeq  \frac{v}{\sqrt{2}} , \ A(\rho) \simeq \frac{n}{q} \frac{1}{\rho} 
\Longrightarrow  
 F(\rho)  \simeq  \frac{v}{\sqrt{2}} , \ a(\rho) \simeq 1 
\nonumber\\ &
\Longrightarrow  
 F(\rho)  \simeq  \frac{v}{\sqrt{2}} , \ \rho w(\rho) \simeq 0 
,
\label{bc1}
\end{align}
and the boundary condition at $\rho=0$:
\begin{align}
 F(\rho)  \sim  \rho^\alpha (\alpha > 0), \quad A(\rho) \sim \rho^\alpha (\alpha > 0) .
\label{bc2}
\end{align}
At the origin $\rho=0$ we require the regularity for the gauge field, which is ensured by
\begin{align}
 & 
\ A(\rho) \sim \rho^\alpha \ (\alpha > 0)    
\Longrightarrow  
a(\rho) \sim \rho^\alpha \ (\alpha > 1)    
\nonumber\\ &
\Longrightarrow  
 w(\rho) = 1/\rho - a(\rho)/\rho \sim 1/\rho + O(\rho^\alpha) \ (\alpha > 0). 
\label{reg1}
\end{align}

We find that the solution satisfying (\ref{bc1}) and (\ref{reg1}) is exactly given by
\begin{align}
 w(\rho) = M_{W} K_1(M_{W}\rho) ,
\end{align}
since the asymptotic forms are given for  large $z$  
\begin{equation}
 K_1(z) \to  \sqrt{\frac{\pi}{2}} z^{-1/2} e^{-z} \left[1+ \frac{3}{8} z^{-1} +O(z^{-2}) \right] \ (|z| \gg 1) , 
\end{equation}
and for small $z$ 
\begin{equation}
 K_1(z) \sim  \frac{1}{z} + \frac14 \left( -1 +2\gamma+2\log \frac{z}{2} \right) z + O(z^3)  \ (|z| \ll 1) .
\end{equation}
Thus, the solution of the gauge field $A$ is decomposed into the massive mode $W$ and the residual mode $V$ as
\begin{align}
 A_a =& V_a + W_a , \
\nonumber\\  
V_a =& - \frac{n}{q}  \epsilon_{ab} \frac{x_b}{\rho} \frac{ 1 }{ \rho} 
= - \frac{n}{q} \frac{\epsilon_{ab} x_b}{x_1^2+x_2^2} ,   
\nonumber\\  
W_a=& \frac{n}{q} \epsilon_{ab} \frac{x_b}{\rho}   w(\rho) 
= \frac{n}{q} \frac{\epsilon_{ab} x_b}{\sqrt{x_1^2+x_2^2}}  M_{W} K_1(M_{W}\rho) .   
\end{align}
The  decomposed fields $V$ and $W$ are singular at $\rho=0$, but singularities cancel between $V$ and $W$ so that $A$ is regular everywhere.

Notice that despite the non-trivial topology of the vacuum manifold there are no finite energy field configurations with non-zero topological charge in the global theory.



\section{$SU(2)$ gauge-scalar model: comparison of fundamental scalar and the adjoint scalar}

We summarize the formulas for the $SU(2)$ gauge-scalar model complementary to the massive $SU(2)$ Yang-Mills theory to see the differences between the fundamental scalar and the adjoint scalar as follows.

In the adjoint scalar case, it is shown \cite{KKSS15} that the partition function is rewritten 
\begin{align}
 Z_{\rm RF} =&  \int \mathcal{D}\hat{\bm{\phi}}   \mathcal{D} \mathscr{A} \delta(\chi)  \Delta^{\rm red}  e^{iS_{\rm YM}[\mathscr{A}]+iS_{\rm kin}[\mathscr{A},\hat{\bm{\phi}}]} ,
 \nonumber\\
   =& \int \mathcal{D}\hat{\bm{\phi}}  \mathcal{D}c  \mathcal{D}\mathscr{W} {J} \delta(\tilde\chi) \tilde\Delta^{
\rm red}   e^{iS_{\rm YM}[\mathscr{V}+\mathscr{W}]+iS_{\rm m}[\mathscr{W} ]} . 
\label{Z-adj}
\end{align}
We can reproduce the preceding cases by choosing the gauge. 
For instance, the unitary gauge, 
\begin{align}
 & \bm{\phi}^A(x) = v \hat{\bm{\phi}}^A(x) , \ 
\hat{\bm{\phi}}^A(x) \to \delta^{A3} ,
\end{align}
 reproduces  
\begin{align}
 Z_{\rm RF} &\to \int \mathcal{D}A^3  \mathcal{D}A^a \delta\left( \mathscr{D}^\mu[A^3]A_\mu^a  \right) \Delta_{\rm FP} e^{iS_{\rm YM}[\mathscr{A}]+iS_{\rm m}[A^a]} ,
\end{align}
since
\begin{align}
  c_\mu=\mathscr{A}_\mu \cdot \hat{\bm{\phi}} \to A_\mu^3, \  \mathscr{W}_\mu \to   A_\mu^a  
\end{align}
In the limit, the gauge-adjoint scalar model with the radially fixed scalar field is reduced to the Yang-Mills theory with the gauge-fixing term of the Maximal Abelian gauge $\mathscr{D}^\mu[A^3]A_\mu^a=0$ and the associated Faddeev-Popov determinant $\Delta_{\rm FP}$ supplemented with a mass term $S_{\rm m}[A^a]$ for the off-diagonal gluons.

\begin{widetext}

\renewcommand{\arraystretch}{1.3}
 \begin{tabular}{l||c|c}
 \multicolumn{3}{c}{$SU(2)$ gauge-scalar model (``complementary'') to the massive $SU(2)$ Yang-Mills theory}\\
\hline
  $G=SU(2)$   & Fundamental scalar $\Phi$, $\Theta = (\tilde\Phi,	\Phi) \in G$ &  Adjoint scalar $\bm{\phi} \in \mathscr{G}$  \\ 
\hline\hline
  SSB pattern $G \to H$ & complete: $SU(2) \to \{ 1 \}$ & partial: $SU(2) \to U(1)$  \\ 
\hline\hline
  field decomposition &  $\mathscr{A}_\mu=\mathscr{W}_\mu+\mathscr{V}_\mu$ &  $\mathscr{A}_\mu=\mathscr{W}_\mu+\mathscr{V}_\mu$  \\ 
  gauge transformation & $\mathscr{A}_\mu \to U\mathscr{A}_\mu U^\dagger+ig^{-1}U\partial_\mu U^\dagger$ & same as on the left \\ 
\hline
  massive mode $\mathscr{W}_\mu$ & $\mathscr{W}_\mu=-ig^{-1} \hat{\Theta} 	({D}_{\mu}[\mathscr{A}] \hat{\Theta} )^\dagger$ & $\mathscr{W}_\mu=-ig^{-1}[\hat{\bm{\phi}},\mathscr{D}_\mu[\mathscr{A}]\hat{\bm{\phi}}]$  
\\ 
 $\mathscr{W}_\mu = \mathscr{W}_\mu^A T_A \in su(2)$   & $\mathscr{W}_{\mu}^A =  ig^{-1} [\hat{\Phi}^\dagger \sigma_A D_{\mu}[\mathscr{A}] \hat{\Phi}- (D_{\mu}[\mathscr{A}] \hat{\Phi} )^\dagger \sigma_A \hat{\Phi} ]$ &  $\mathscr{W}_{\mu}^A =   g^{-1}  \epsilon^{ABC} \hat\phi^B (\mathscr{D}_\mu[\mathscr{A}]\hat\phi)^C$    
\\
  gauge transformation & $\mathscr{W}_\mu \to U\mathscr{W}_\mu U^\dagger$  & same as on the left \\ 
\hline
  residual mode $\mathscr{V}_\mu$  & $\mathscr{V}_\mu=ig^{-1} \hat{\Theta} \partial_{\mu}  \hat{\Theta} ^\dagger$ &  $\mathscr{V}_\mu=c_\mu \hat{\bm{\phi}}+ig^{-1}[\hat{\bm{\phi}},\partial_\mu \hat{\bm{\phi}}]$ 
\\ 
 $\mathscr{V}_\mu = \mathscr{V}_\mu^A T_A \in su(2)$    & $\mathscr{V}_{\mu}^A =  -ig^{-1} [\hat{\Phi}^\dagger \sigma_A \partial_\mu \hat{\Phi}- \partial_\mu \hat{\Phi} ^\dagger \sigma_A \hat{\Phi} ]$ &  $ c_\mu=\mathscr{A}_\mu \cdot \hat{\bm{\phi}}$ 
\\
  gauge transformation & $\mathscr{V}_\mu \to U\mathscr{V}_\mu U^\dagger+ig^{-1}U\partial_\mu U^\dagger$ & same as on the left \\ 
\hline
Defining equation  & $ {D}_\mu[\mathscr{V}] \hat{\Phi}=0$, $ {D}_\mu[\mathscr{V}] \hat{\Theta}=0$ & $\mathscr{D}_\mu[\mathscr{V}] \hat{\bm{\phi}}=0$ \\
   & ($ \mathscr{W} \cdot \hat{\Phi}\not=0$)  & $ \mathscr{W}_\mu \cdot \hat{\bm{\phi}} =0$ \\
\hline
  field equation 1 & ${\rm tr} \left(\Theta ^{\dagger}\Theta  - \frac{1}{2}v^2 \bm{1} \right) /{\rm tr}(\bm{1}) = 0$ & $\bm{\phi}  \cdot \bm{\phi}  - v^2 = 0$ \\ 
  field equation 2 & $-  {D}_{\mu}[\mathscr{A}] {D}^{\mu}[\mathscr{A}] \Theta  + \Theta u  = 0$ & $-  \mathscr{D}^{\mu}[\mathscr{A}] \mathscr{D}_{\mu}[\mathscr{A}] \bm{\phi}  
   +  2u \bm{\phi} =0$ \\ 
  field equation 3 & $\mathscr{D}^\nu [\mathscr{A}] \mathscr{F}_{\nu\mu}[\mathscr{A}]  +   M_W^2 \mathscr{W}_\mu = 0$ & $\mathscr{D}^\nu [\mathscr{A}] \mathscr{F}_{\nu\mu}[\mathscr{A}]  +   M_W^2 \mathscr{W}_\mu = 0$ \\ 
\hline
  reduction condition $\chi$ & $\mathscr{D}^\mu[\mathscr{V}]\mathscr{W}_\mu=0$ & $\mathscr{D}^\mu[\mathscr{V}]\mathscr{W}_\mu=0$ \\ 
\hline
  color direction field  &  $\bm{n}   
=  \hat\Theta \sigma_3 \hat\Theta^{\dagger} $ & $\bm{n} =\hat{\bm{\phi}} $  \\ 
  $\bm{n}=n^A \sigma_A $   &  $  n^A  
= - \hat{\Phi}^\dagger \sigma_A \hat{\Phi}$ & $n^A=\hat{\phi}^A$  \\ 
 \end{tabular}
\vskip 0.5cm

\end{widetext}



\begin{thebibliography}{99}
\bibitem{Kondo16}
K.-I. Kondo, 
Phys. Lett. B{\bf 762},  219--224 (2016).
arXiv:1606.06194 [hep-th]  
\\
K.-I. Kondo, 
EPJ Web Conf. 137 (2017) 03009. 
arXiv:1612.05933 [hep-th]





\bibitem{Higgs1}
P.W. Higgs,
Phys. Lett. \textbf{12},  132
 (1964).
\\
P.W. Higgs,
Phys. Rev. Lett.  \textbf{13},  508
 (1964).


\bibitem{Higgs2}
F. Englert and R. Brout,
Phys. Rev. Lett. \textbf{13}, 321
 (1964). 


\bibitem{Higgs3}
G.S. Guralnik, C.R. Hagen, and T.W.B. Kibble,
Phys. Rev. Lett. \textbf{13}, 585 (1964). 


\bibitem{NJL61}
Y. Nambu and G. Jona-Lasinio, 
Phys.Rev. \textbf{112}, 345
(1961).


\bibitem{Goldstone61}
J. Goldstone, 
Nuovo Cimento \textbf{19}, 154--164 (1961).
\\
J. Goldstone, A. Salam and S. Weinberg, 
Phys. Rev. \textbf{127}, 965
 (1962).


 

\bibitem{Elitzur75}
S. Elitzur, 
Phys. Rev. D\textbf{12}, 3978
 (1975).  
\\
G. F. De Angelis, D. de Falco, and F. Guerra, 
Phys.Rev. D\textbf{17}, 1624
 (1978).  
















\bibitem{FS79}
E. Fradkin and S. Shenker,
Phys. Rev. D\textbf{19}, 3682
  (1979). 


\bibitem{OS78}
K. Osterwalder and E. Seiler,
Ann. Phys. \textbf{110}, 440
 (1978).
\\
E. Seiler, 
Lect. Notes Phys. \textbf{159}, 1
 (1982).  


\bibitem{lattice-gauge-scalar-fund}
M. Creutz, L. Jacob, C. Rebbi,
Phys. Rept. {\bf 95}, 201--282 (1983). 


K. Langfeld,
hep-lat/0212032 


C. Bonati, G. Cossu, M. D'Elia, A. Di Giacomo,
Nucl. Phys. B{\bf 828}, 390--403 (2010). 
arXiv:0911.1721 [hep-lat]


\bibitem{lattice-gauge-scalar-adj}
C.B. Lang, C. Rebbi, M. Virasoro,
Phys. Lett. {\bf 104}B, 294 (1981). 

R.C. Brower, D.A. Kessler, T. Schalk, H. Levine, and M. Nauenberg,
Phys. Rev. D\textbf{25}, 3319
 (1982).  


I-Hsiu Lee and J. Shigemitsu,
Nucl. Phys. B\textbf{263},  280--294 (1986).


S. Nadkarni, 
Nucl. Phys. B{\bf 334},  559--579 (1990). 

A. Hart, O. Philipsen, J.D. Stack, M. Teper,
Phys. Lett. B{\bf 396} , 217--224 (1997). 
hep-lat/9612021


J. Greensite, S. Olejnik, D. Zwanziger,
Phys. Rev. D\textbf{69}, 074506 (2004).  
hep-lat/0401003 


\bibitem{gauge-scalar-rad-var}
T. Munehisa and Y. Munehisa, 
Nucl. Phys. B{\bf 215}, 508--526 (1983), Erratum: Nucl. Phys. B{\bf 218}, 545--545 (1983). 


  
Y. Munehis, 
Phys.Rev. D{\bf 30},  1310 (1984). 


Y. Munehisa,
Phys.Rev. D{\bf 31},  1522 (1985). 


Y. Munehisa,
Phys.Lett. {\bf 155}B,  159--162 (1985). 


T. Munehisa and Y. Munehisa, 
Z.Phys. C{\bf 32},  531 (1986). 


Y. Munehisa, 
Mod.Phys.Lett. A{\bf 3},  23--31 (1988). 


K.-I. Kondo,
Nucl. Phys. B {\bf 295} [FS21], 93--104 (1988).


J. Jersak,
Lattice Studies Of The Higgs System,  
HLRZ-89-45, DESY-89-115 

J. Jersak,
Lattice Higgs Models,  
Published in NATO Sci.Ser.B 140, 133--169  (1985).
PITHA-85/25 



\bibitem{FMS80}
J. Fr\"ohlich, G. Morchio, and F. Strocchi,
Phys. Lett. B\textbf{97}, 249
 (1980).  
Nucl. Phys. B\textbf{190}, 553
  (1981). 


\bibitem{tHooft80}
G. 't Hooft,
Which Topological Features of a Gauge Theory Can Be Responsible for Permanent Confinement? 
Lecture given at Cargese Summer Inst.1979, 
NATO Sci.Ser.B \textbf{59}, 117 (1980). 


\bibitem{Maas17}
A. Maas,
arXiv:1712.04721 [hep-ph] 


\bibitem{soliton}
R. Rajaraman,
``Solitons and Instantons: An Introduction to Solitons and Instantons in Quantum Field Theory''
(North-Holland, Amsterdam, 1987).
\\
A. Vilenkin and E.P.S. Shellard,
``Cosmic Strings and Other Topological Defects"
(Cambridge University Press, 2000).
\\
N. Manton and P. Sutcliffe,
``Topological Solitons"
(Cambridge University Press, 2004).
\\
Ya. Shnir,
``Magnetic Monopoles''
(Springer, 2005).
\\
E.J. Weinberg,
``Classical Solutions in Quantum Field Theory''
(Cambridge University Press, 2012).


\bibitem{Polyakov77b}
  A.M. Polyakov,
Nucl. Phys. B\textbf{120}, 429
 (1977).


\bibitem{tHP74}
G. 't Hooft,
Nucl. Phys. B\textbf{79}, 276
 (1974).
\\
A. M. Polyakov,
JETP Lett. \textbf{20}, 194
 (1974).


\bibitem{Cho80}
  Y.M. Cho,
Phys. Rev. D{\bf 21}, 1080
 (1980).


\bibitem{DG79}
  Y.S. Duan and M.L. Ge, 
Sinica Sci., {\bf 11}, 1072
(1979). 


\bibitem{FN99} 
L. Faddeev and A. Niemi,
Phys.Rev. Lett.{\bf 82}, 1624
(1999).


\bibitem{KMS06}
  K.-I. Kondo, T. Murakami and T. Shinohara,
Prog. Theor. Phys. {\bf 115}, 201
(2006). 


\bibitem{KMS05}
  K.-I. Kondo, T. Murakami and T. Shinohara,
Eur. Phys. J. C{\bf 42}, 475
(2005).


\bibitem{Kondo06}
K.-I. Kondo,
Phys. Rev. D{\bf 74}, 125003 (2006). 


\bibitem{KSM08}
K.-I. Kondo, T. Shinohara and T. Murakami,
Prog. Theor. Phys. {\bf 120},  1
(2008).


\bibitem{KKSS15}
K.-I. Kondo, S. Kato, A. Shibata and T. Shinohara,
Phys. Rept. \textbf{579}, 1
 (2015).
arXiv:1409.1599 [hep-th].


\bibitem{NMWK18}
S. Nishino, R. Matsudo, M. Warschinke, and K.-I. Kondo, 
arXiv:1803.04339 [hep-th]
 
 
\bibitem{Kondo97}
K.-I. Kondo,
Phys. Rev. D{\bf 57}, 7467
(1998). 


\bibitem{Stueckelberg38}
E.C.G. Stueckelberg,
Helv. Phys. Acta {\bf 11}, 225
 (1938).  


\bibitem{KG67}
T. Kunimasa and T. Goto,  
Prog. Theor. Phys. {\bf 37}, 452
 (1967). 
\\
T. Fukuda, M. Monda, M. Takeda and Kan-ichi Yokoyama,
Prog. Theor. Phys. {\bf 66}, 1827
 (1981).


\bibitem{SF70}
A.A. Slavnov and L.D. Faddeev, 
Theor. Math. Phys. {\bf 3},  312
 (1970) [Teor. Mat. Fiz. {\bf 3}, 18
 (1970)]. 
\\
A.A. Slavnov,
Theor. Math. Phys. {\bf 10},  201
 (1972) [Teor. Mat. Fiz. {\bf 10}, 305
 (1972)]. 


\bibitem{Cornwall74}
J. M. Cornwall,
Phys. Rev. D{\bf 10}, 500
 (1974). 
\\
J. M. Cornwall,
Nucl. Phys. B{\bf 157}, 392
 (1979).  

\bibitem{Cornwall82}
J. M. Cornwall,
Phys. Rev. D{\bf 26}, 1453
 (1982). 


\bibitem{DT86}
R. Delbourgo and G. Thompson,
Phys. Rev. Lett. A{\bf 57}, 2610
 (1986).


\bibitem{DTT88}
R. Delbourgo, S. Twisk and G. Thompson,
Int. J. Mod. Phys. A{\bf 3}, 435
 (1988).


\bibitem{RRA04}
H. Ruegg and M. Ruiz-Altaba,
[hep-th/0304245],
Int. J. Mod. Phys. A{\bf 19},  3265
 (2004).


\bibitem{decoupling-lattice}
I.L. Bogolubsky, E.-M. Ilgenfritz, M. M\"uller-Preussker and A. Sternbeck,
Phys. Lett. B{\bf 676} (2009) 69--73.  
arXiv:0901.0736 [hep-lat]  


 A. Cucchieri and T. Mendes,
Phys. Rev.D{\bf 78}, 094503 (2008).
arXiv:0804.2371[hep-lat] 
\\
Phys. Rev. Lett. {\bf 100} (2008) 241601.  
arXiv:0712.3517 [hep-lat]  



O. Oliveira and P.J. Silva,
Phys. Rev. D{\bf 86} (2012) 114513. 
arXiv:1207.3029 [hep-lat] 


A.G. Duarte, O. Oliveira, and P.J. Silva, 
Phys. Rev. D{\bf 94} (2016) no.1, 014502. 
arXiv:1605.00594 [hep-lat]



\bibitem{decoupling-analytical}
 Ph. Boucaud, J.P. Leroy, A. Le Yaouanc, J. Micheli, O. Pene and J. Rodriguez-Quintero, 
JHEP 0806, 099 (2008).
hep-ph/0803.2161,


 A.C. Aguilar, D. Binosi and J. Papavassiliou,
Phys. Rev. D{\bf 78}, 025010 (2008).
arXiv:0802.1870 [hep-ph],


C.S. Fischer, A. Maas and J.M. Pawlowski, 
Annals Phys. {\bf 324}, 2408
 (2009). 
arXiv:0810.1987 [hep-ph] 


J. Braun, H. Gies, J.M. Pawlowski, 
 Phys. Lett. B{\bf 684}, 262
 (2010). 
arXiv:0708.2413 [hep-th]. 


M. Tissier and N. Wschebor, 
Phys.Rev. D{\bf 82}, 101701 (2010). 
arXiv:1004.1607 [hep-ph]. 
\\
M. Tissier and N. Wschebor, 
Phys.Rev. D{\bf 84}, 045018  (2011).
arXiv:1105.2475 [hep-th],


K.-I. Kondo, 
Phys. Rev. D{\bf 84},  061702 (2011). 
arXiv:1103.3829 [hep-th]  


M.A.L. Capri, D. Dudal, A.J. Gomez, M.S. Guimaraes, I.F. Justo, S.P. Sorella, and D. Vercauteren,
Phys.Rev. D{\bf 88} (2013) 085022. 
arXiv:1212.1003 [hep-th] 

M.A.L. Capri, D. Dudal, A.J. Gomez, M.S. Guimaraes, I.F. Justo, and S.P. Sorella,
Eur.Phys.J. C{\bf 73} (2013) no.3, 2346. 
arXiv:1210.4734 [hep-th] 




\bibitem{QCD-TNT09}
PoS QCD-TNT09, 
Proceedings of QCD Green's Functions, Confinement, and Phenomenology (QCD-TNT), Trento, Italy, 7-11 Sep 2009, 
Edited by J.M. Cornwall, D. Binosi, J. Papavassiliou, A.C. Aguilar. 
http://pos.sissa.it/cgi-bin/reader/conf.cgi?confid=87, 


\bibitem{Ghent-QCD}
Proceedings of The many faces of QCD, Ghent, Belgium, 1-5 Nov 2010.
http://sites.google.com/site/facingqcd/


\bibitem{Bowmanetal07}
P.O. Bowman,  U.M. Heller,  D.B. Leinweber, M.B. Parappilly, A. Sternbeck, L. von Smekal, A.G. Williams and  Jian-bo Zhang,  
Phys. Rev.D{\bf 76}, 094505 (2007). 
hep-lat/0703022. 


\bibitem{Kondo-etal18}
K.-I. Kondo, Y. Suda, M. Ohuchi, R. Matsudo, and Y. Hayashi, 
Preprint: CHIBA-EP-231, in preparation. 


\bibitem{KO79}
T. Kugo and I. Ojima,  
Prog. Theor. Phys. Suppl. {\bf 66}, 1--130  (1979). 


\bibitem{CF76b}
G. Curci and R. Ferrari, 
Nuovo Cim. A{\bf 35}, 1
 (1976), Erratum-ibid. A{\bf 47}, 555 (1978). 


\bibitem{Kondo13}
K.-I. Kondo,
Phys. Rev. D\textbf{87}, 025008 (2013).  
e-Print: arXiv:1208.3521 [hep-th]  

K.-I. Kondo, K. Suzuki, H. Fukamachi, S. Nishino, and T. Shinohara,
Phys. Rev. D\textbf{87}, 025017 (2013).  
e-Print: arXiv:1209.3994 [hep-th]   

 
\end{thebibliography}
\end{document}